\documentclass[epj,nopacs]{svjour}
\pdfoutput=1
\usepackage{amsmath}
\usepackage{amsfonts}
\usepackage{amssymb}
\usepackage{color}
\usepackage{graphicx}
\usepackage{mathrsfs}
\usepackage{graphics}
\usepackage{ulem}
\usepackage{enumerate}% Remove option referee for final version
\begin{document}
%

%\vspace*{-3.5cm}
% \begin{flushright}
% {\small
% CCTP-2015-01  \\  CCQCN-2015-59 
% }
%\end{flushright}

\title{Phase transitions in a holographic s+p model with back-reaction}
%\subtitle{Do you have a subtitle?\\ If so, write it here}
\author{Zhang-Yu Nie\inst{1,2,3}\thanks{e-mail:niezy@itp.ac.cn} \and Rong-Gen Cai\inst{2}\thanks{e-mail:cairg@itp.ac.cn} \and Xin Gao\inst{4}\thanks{e-mail:xingao@vt.edu} \and Li Li\inst{5}\thanks{e-mail:lili@physics.uoc.gr} \and Hui Zeng\inst{1,2}\thanks{e-mail:zenghui@kmust.edu.cn}
}
\mail {Zhang-Yu Nie}
\institute{Kunming University of Science and Technology, Kunming 650500, China \and 
State Key Laboratory of Theoretical Physics, Institute of Theoretical Physics, Chinese Academy of Sciences, P.O.Box 2735, Beijing 100190, China \and
INPAC, Department of Physics, and Shanghai Key Laboratory of Particle Physics and Cosmology, Shanghai Jiao Tong University, Shanghai 200240, China \and
Department of Physics, Robeson Hall, 0435, Virginia Tech 850 West Campus Drive, Blacksburg, VA 24061, USA \and
Crete Center for Theoretical Physics, Department of Physics, University of Crete, 71003 Heraklion, Greece}
%
%\date{Received: date / Revised version: date}
\date{CCTP-2015-01  \\  CCQCN-2015-59 }
% The correct dates will be entered by Springer
%
\abstract{
In a previous paper (arXiv:1309.2204, JHEP 1311 (2013) 087), we present a holographic s+p superconductor model with a scalar triplet charged under an SU(2) gauge field in the bulk. We also study the competition and coexistence of the s-wave and p-wave orders in the probe limit. In this work we continue to study the model by considering the full back-reaction The model shows a rich phase structure and various condensate behaviors such as the ``n-type'' and ``u-type'' ones, which are also known as reentrant phase transitions in condensed matter physics.  The phase transitions to the p-wave phase or s+p coexisting phase become first order in strong back-reaction cases. In these first order phase transitions,  the free energy curve always forms a swallow tail shape, in which the unstable s+p solution can also play an important role. The phase diagrams of this model are given in terms of the dimension of the scalar order and the temperature in the cases of eight different values of the back reaction parameter, which show that the region for the s+p coexisting phase is enlarged with a small or medium back reaction parameter, but is reduced in the strong back-reaction cases.
%
%\PACS{
 %     {PACS-key}{discribing text of that key}   \and
 %     {PACS-key}{discribing text of that key}  CCTP-2015-01   CCQCN-2015-59
 %   } % end of PACS codes
} %end of abstract

\maketitle

\section{Introduction}
\label{sec:intro}

Over the past years,the AdS/CFT correspondence~\cite{Maldacena:1997re,Gubser:1998bc,Witten:1998qj} has attracted  a lot of attention as a tool to study strongly coupled systems. One of the most successful applications of the correspondence is the so-called holographic superconductor which is first proposed in Refs.~\cite{Gubser:2008px,Hartnoll:2008vx}. In this model, the condensation of a charged field in asymptotically AdS spacetime can trigger the superconducting phase transition in the dual field theory. According to the different property, especially the symmetry of the condensed fields, the holographic model has been extended to the p-wave and d-wave superconductor models~\cite{Gubser:2008wv,Cai:2013aca,Chen:2010mk,Kim:2013oba,Zeng:2010zn}.

Recently the competition and coexistence between different order parameters have been studied in some holographic framework. In Refs.~\cite{Basu:2010fa,Cai:2013wma,Chaturvedi:2014dga}, the authors studied the system with two s-wave orders. Refs.~\cite{Nie:2013sda,Amado:2013lia,Momeni:2013bca} presented the results of the holographic systems with s-wave and p-wave orders simultaneously. The holographic systems with one s-wave order and one d-wave order have also been investigated in Refs.~\cite{Nishida:2014lta,Li:2014wca}. More works on holographic study of the multi-order systems can be found in Refs.~\cite{Huang:2011ac,DG,Krikun:2012yj,Donos:2012yu,Musso:2013ija,AAJL,Nitti:2013xaa,Liu:2013yaa,Amoretti:2013oia,Donos:2013woa,Cai:2014dza}. These studies might shed light to the understanding of some universal properties of the competition and coexistence between different orders in real materials such as high $T_c$ cuprates and $^3\text{He}$ superfluid.

In a previous work~\cite{Nie:2013sda}, we built a holographic model with a scalar triplet charged under an SU(2) gauge field in the bulk and studied the competition and coexistence between the s-wave and p-wave superconducting order parameters. The results showed that the system could be in an s-wave phase at higher temperatures and then transfers to a p-wave phase at lower temperatures via an s+p coexisting phase. In that work, we worked in the probe limit and found that the temperature region for the s+p coexisting phase is very narrow. The free energy curves of these phases show that the s+p phase is always stable.

Our previous study in the probe limit has revealed a typical phase transition behavior among the s-wave phase, the p-wave phase and the s+p phase. However, the results in the probe limit can only be meaningful when the charge of the s-wave and p-wave orders is vary large. If one wants to uncover the complete phase diagram of this s+p model with finite charge, we must go beyond the probe limit and take into account the back reaction of the matter fields on the metric. Moreover, note that in previous studies, the back reacted s+s model has already shown new phase transition behaviors which do not exist in the probe limit~\cite{Basu:2010fa,Cai:2013wma}. In the s+p model, the condensate of p-wave order would break the spatial rotational symmetry spontaneously, which is obviously different from the s+s case. Thus we can expect to discover new phase transition behaviors in the back reacted s+p model with an anisotropic p-wave order. Finally, the holographic p-wave superconductors with back-reaction have already been well studied~\cite{Arias:2012py,Gubser:2010dm,Herzog:2014tpa,Ammon:2009xh}, it was found that the phase transition controlled by the p-wave order becomes first order when the back reaction is strong enough~\cite{Gubser:2010dm,Herzog:2014tpa,Ammon:2009xh}. Therefore, it is necessary to study whether there is still an s+p phase and what would happen to the s+p phase when the order of phase transition changes. In summary, it is interesting and important to take into account back-reaction in the s+p model to get the rich phenomena of phase transitions, and to reach a better understanding on systems with competing orders.

In this paper, we turn on the back reaction of matter fields on the background geometry in our holographic s+p model~\cite{Nie:2013sda}. More precisely, the strength of the back reaction is controlled by a parameter $b$. For simplicity, we consider eight values of back reaction parameter $b$ varying from $0.1$ to $0.8$, and build corresponding  phase diagrams in terms of the dimension of the scalar order and the temperature. We also give an alternative setup of the holographic s+p system, which would allow us to vary the charge and dimension of the p-wave order.

This article is organized as follows. In the next section, we briefly introduce our holographic model and the definition for the parameters. In section \ref{sect:FreeE} we calculate the free energy of the system. In section \ref{sect:PT}, we show the condensation behavior of some interesting phase transitions with different back reaction strength. We also give the free energy behavior for some special cases with unstable s+p solution to illustrate the important role played by the unstable s+p solution. In section \ref{sect:PD}, we construct the phase diagrams with eight different values of back reaction parameter. In section \ref{sect:newSetup}, we give  another setup of a holographic s+p model, from which with some special choice of model parameters, we can get the same equations of motions and thus the same results of phase transitions. We summarize our results and make some outlooks in section \ref{sect:conclusion}.

\section{Holographic model of an s+p superconductor}
\label{sect:setup}

The action of our holographic s+p superconductor model takes the following form~\cite{Nie:2013sda}
\begin{eqnarray}
S  &=&S_G+S_M,\\
S_G&=&\frac{1}{2 \kappa_g ^2}\int d^{d+1}x \sqrt{-g} (R-2\Lambda),\\
S_M&=&\frac{1}{g_c^2}\int d^{d+1}x \sqrt{-g}(-D_\mu \Psi^{a} D^\mu \Psi^a \nonumber\\
&&-\frac{1}{4}F^a_{\mu\nu}F^{a\mu\nu}-m^2 \Psi^a\Psi^a). \label{Smatter}
\end{eqnarray}
Here $\Psi^a$ is an $SU(2)$ scalar triplet and $A_\mu^a$ is the $SU(2)$ gauge field with $F^{a}_{\mu\nu}$ the field strength. $a=(1,2,3)$ is the index of the generators of SU(2) algebra. From this action, we can get the equations of motion for the matter fields
\begin{eqnarray}
g^{\mu\nu} D_\mu D_\nu \Psi^a -m^2 \Psi^a &=&0,\\
g^{\rho\nu} D_\rho F_{\nu\mu}^a - 2 \varepsilon^{abc} \Psi^b D_\mu \Psi^c &=&0,
\end{eqnarray}
and for the gravitational fields
\begin{equation}
R_{\mu\nu} -\frac{1}{2}(R-2\Lambda) g_{\mu\nu} = b^2 \mathcal{T}_{\mu\nu},
\end{equation}
where $b= \kappa_g/g_c$ characterizes the strength of the back reaction of the matter field on the background geometry, and $\mathcal{T}_{\mu\nu}$ is 
the stress-energy tensor of the matter sector
\begin{eqnarray}
\mathcal{T}_{\mu\nu} = &(-D_\mu \Psi^a D^\mu \Psi^a -\frac{1}{4}F_{\mu\nu}^a F^{a\mu\nu} -m^2 \Psi^a \Psi^a) g_{\mu\nu} \nonumber\\
&-2(-D_\mu \Psi^a D_\nu \Psi^a -\frac{1}{2}F_{\mu\rho}^a F_\nu^{a\rho}).
\end{eqnarray}
Here $D_\mu$ is the covariant derivative with SU(2) connection on the curved spacetime background and $\varepsilon^{abc}$ is antisymmetric with $\varepsilon^{123}=1$. The action of $D_\mu$ on a spacetime tensor $T_{\mu_1...\mu_n}^a$ with one group index $a$ is given by
\begin{equation}
D_\mu T_{\mu_1...\mu_n}^a=\nabla_\mu T_{\mu_1...\mu_n}^a + \varepsilon^{abc} A_\mu^b T_{\mu_1...\mu_n}^c.
\end{equation}

We will work in $d=3$ case as in Ref.~\cite{Nie:2013sda}. In order to get both the s-wave and p-wave superconducting phases in this model, we can choose the $A^1_\mu$ field as the electro-magnetic $U(1)$ gauge field, and $\Psi^3$, $A^3_x$ as the scalar and vector order parameters, respectively. Thus the ansatz for the matter fields is
\begin{eqnarray}\label{matterAnsatz}
\Psi^3=\Psi_3(r),~A^1_t=\phi(r),~A^3_x=\Psi_x(r),
\end{eqnarray}
and all the other components of the scalar and gauge field are set to be zero.

The black hole geometry  compatible with the ansatz  of matter fields  (\ref{matterAnsatz}) is~\cite{Ammon:2009xh,Manvelyan:2008sv,Cai:2010zm}
\begin{eqnarray}
ds^2=&-N(r)\sigma(r)^2dt^2 + \frac{1}{N(r)}dr^2 \nonumber\\
&+r^2f(r)^{-2}dx^2+r^2f(r)^2dy^2,
\end{eqnarray}
with
\begin{equation}
N(r)=\frac{r^2}{L^2}(1-\frac{2 M(r)}{r^3}),
\end{equation}
where $L$ is the AdS radius. 
The function $M(r)$ is related to the mass and charge of the black brane. The horizon of the black brane is located at $r=r_h$ where $M(r_h)=r_h^3/2,$ and the temperature is
\begin{equation}\label{TemperatureE}
T=\frac{N'(r_h) \sigma(r_h)}{4\pi}.
\end{equation}
The full equations of motion for this system under above ansatz  read
\begin{eqnarray}\label{EoMs1}
M'(r)&=& \nonumber
\frac{b^2 L^2 f(r)^2 \Psi_x(r)^2 \phi (r)^2}{4 N(r) \sigma (r)^2}
+\frac{1}{4} b^2 L^2 f(r)^2 N(r) \Psi_x'(r)^2\\&& \nonumber
+\frac{1}{2} b^2 L^2 m^2 r^2 \Psi_3(r)^2
+\frac{b^2 L^2r^2 \Psi_3(r)^2 \phi (r)^2}{2 N(r) \sigma (r)^2}\\&&\nonumber
+\frac{1}{2} b^2 L^2 r^2 N(r) \Psi_3'(r)^2+\frac{b^2 L^2 r^2 \phi '(r)^2}{4 \sigma (r)^2}\\&&
+\frac{r^2 L^2 N(r) f'(r)^2}{2 f(r)^2},
\\
\label{EoMs2} \sigma '(r)&=&\frac{b^2 f(r)^2 \Psi_x(r)^2 \phi (r)^2}{2 r N(r)^2 \sigma (r)}+\frac{b^2 f(r)^2 \sigma (r) \Psi_x'(r)^2}{2 r}\nonumber\\&&
+\frac{b^2 r \Psi_3(r)^2 \phi (r)^2}{N(r)^2 \sigma (r)}
+b^2 r \sigma (r) \Psi_3'(r)^2\nonumber\\&&
+\frac{r \sigma (r) f'(r)^2}{f(r)^2},
\\
\label{EoMs3} f''(r)&=&-\frac{b^2 f(r)^3 \Psi_x(r)^2 \phi (r)^2}{2 r^2 N(r)^2 \sigma (r)^2}+\frac{b^2 f(r)^3 \Psi_x'(r)^2}{2 r^2}+\frac{f'(r)^2}{f(r)}\nonumber\\&&
-\frac{f'(r) N'(r)}{N(r)}-\frac{f'(r) \sigma '(r)}{\sigma (r)}-\frac{2 f'(r)}{r},\\
\label{EoMs4} \phi ''(r)&=&\frac{f(r)^2 \Psi_x(r)^2 \phi (r)}{r^2 N(r)}+\frac{2 \Psi_3(r)^2 \phi (r)}{N(r)}\nonumber\\&&
+\frac{\sigma '(r) \phi '(r)}{\sigma (r)}-\frac{2 \phi '(r)}{r},\\
\label{EoMs5} \Psi_x''(r)&=&-\frac{2 f'(r) \Psi_x'(r)}{f(r)}-\frac{N'(r) \Psi_x'(r)}{N(r)}\nonumber\\&&
-\frac{\Psi_x(r) \phi (r)^2}{N(r)^2 \sigma (r)^2}-\frac{\sigma '(r) \Psi_x'(r)}{\sigma (r)},\\
\label{EoMs6} \Psi_3''(r)&=&\frac{m^2 \Psi_3(r)}{N(r)}-\frac{N'(r) \Psi_3'(r)}{N(r)}-\frac{\Psi_3(r) \phi (r)^2}{N(r)^2 \sigma (r)^2}\nonumber\\&&
-\frac{\sigma '(r) \Psi_3'(r)}{\sigma (r)}-\frac{2 \Psi_3'(r)}{r}.
\end{eqnarray}
Note that in above equations, there exist  four sets of scaling symmetries as follows
\begin{eqnarray}
(1)&N \rightarrow \lambda^2 N~,~L\rightarrow \lambda^{-1} L~,~b \rightarrow \lambda^{-1} b~, ~\sigma \rightarrow \lambda^{-1} \sigma~,\nonumber\\&
\phi \rightarrow \lambda \phi~,~\Psi_x \rightarrow \lambda \Psi_x~,~\Psi_3 \rightarrow \lambda \Psi_3~,~m\rightarrow \lambda m~;
\\(2)& \phi \rightarrow \lambda \phi~,~\Psi_x \rightarrow \lambda \Psi_x~,~N \rightarrow \lambda^2 N~,\nonumber\\&
~M \rightarrow \lambda^3 M~,~r \rightarrow \lambda r~;
\\(3)& \phi \rightarrow \lambda \phi~,~\sigma \rightarrow \lambda \sigma~; \label{scaling3}
\\(4)& \Psi_x \rightarrow \lambda \Psi_x~,~f \rightarrow \lambda^{-1} f~. \label{scaling4}
\end{eqnarray}
These scaling symmetries are very useful in the numerical calculations. For example, we can use the first scaling symmetry to set $L=1$ and the second one to set $r_h=1$. After we get the numerical solutions, we can use these scaling symmetries again to recover $L$ and $r_h$ to any value. The last two scaling symmetries will be used to scale the solutions to asymptotically AdS type with $\lim\limits_{r\to\infty} \sigma(r) \rightarrow1$ and $\lim\limits_{r\to\infty}f(r)\rightarrow 1$. 

In order to solve the equations of motion numerically, we need to specify the boundary conditions both on the horizon $r=r_h$ and on the boundary $r=\infty$. Without loss of generality, we set $L=1$ in the rest of the paper. Then near the horizon, the functions can be expanded as
\begin{eqnarray}
M(r) &=& \frac{r_h^3}{2}+M_{h1} (r-r_h) + ...~,\\
\sigma(r) &=& \sigma_{h0}+\sigma_{h1}(r-r_h)+...~,\\
f(r) &=& f_{h0} + f_{h1}(r-r_h)+...~,\\
\phi(r) &=& \phi_{h1}(r-r_h)+\phi_{h2}(r-r_h)^2+...~,\\
\Psi_x(r) &=& \Psi_{xh0}+\Psi_{xh1}(r-r_h)+...~,\\
\Psi_3(r) &=& \Psi_{3h0}+\Psi_{3h1}(r-r_h)+... ~.
\end{eqnarray}
One can check that only the coefficients $\{\sigma_{h0},f_{h0},\phi_{h1},\Psi_{xh0}, \Psi_{3h0} \}$ are independent. That means when we work at some fixed value of $b$ and $m^2$, we can get a group of solutions with these five independent parameters. However, we should further consider some constraints from the boundary $r=\infty$. Then the degrees of freedom of the solutions would be reduced.

We can expand the functions near the  AdS boundary as
\begin{eqnarray}
M(r)= &M_{b0}+ \frac{M_{b1}}{r} + ...~,
&\sigma(r)=\sigma_{b0}+ \frac{\sigma_{b3}}{r^3}+...~,\nonumber\\
f(r)= &f_{b0} +  \frac{f_{b3}}{r^3}+...~,
&\Psi_3(r)=\frac{\Psi_{3S}  }{r^{3-\Delta}}+ \frac{\Psi_{3E} }{r^{\Delta}}+... ~,
\nonumber\\
\Psi_x(r)= &\Psi_{xb0}+  \frac{\Psi_{xb1}}{r}+...~,
&\phi(r)=\mu - \frac{\rho }{r}+...~,
\end{eqnarray}
where $\Delta=(3+\sqrt{9+4m^2})/2$ is the scaling dimension of the scalar order. Note that there is no mass term for the SU(2) fields, thus the dimension for the p-wave order is fixed to be $\Delta_p=d-1=2$.

In order to make the boundary geometry to be asymptotically AdS(a-AdS), we should have $\sigma_{b0}=f_{b0}=1$. These two conditions could be easily satisfied by a scale transformation from any known solution by using the last two scaling symmetries (\ref{scaling3},\ref{scaling4}).

In the asymptotically  AdS spacetime, we can use the AdS/CFT correspondence to get information of the dual field theory. The AdS/CFT dictionary tells us that $\mu$ and $\rho$ are related to the chemical potential and charge density respectively, while $\Psi_{3S},\Psi_{xb0}$ are related to the sources and $\Psi_{3E},\Psi_{xb1}$ are related to the expectation values of the dual s-wave and p-wave operators. To study the phases where the symmetry is spontaneously broken, we further impose the source free condition with $\Psi_{xb0}=\Psi_{3S}=0$. There is an alternative quantization with the expression of the source and expectation interpretation exchanged~\cite{Marolf:2006nd,Klebanov:1999tb,Ren:2010ha,Gao:2012yw}. But  here we only focus on the standard quantization case.

With these conditions from the boundary, we can count the degrees of freedom left for the equations. Besides the two parameters $b$ and $\Delta$ from the model, we have five free parameters at horizon to fix the solution. The solution we need should satisfy four conditions on the boundary. Thus the solutions satisfying the four boundary conditions would have only one free parameter besides $b$ and $\Delta$. In this paper, we would choose this free parameter as $T$ while fixing the value of $\mu$. Thus we work in the grand canonical ensemble, and we can plot our numerical results with respect to the dimensionless parameter $T/\mu$. This is also convenient for comparing the results in this paper to that in the probe limit in Ref.~\cite{Nie:2013sda}.

\section{Free Energy}\label{sect:FreeE}
In our model, there left one degree of freedom denoted by $T$ for the solutions, once we have fixed the values of $b$ and $\Delta$.  The model includes four different solutions, which are the AdS Reissner-Nordstr\"om (RN)  black brane solution describing the normal conductor phase, the s-wave superconductivity solution, the p-wave superconductivity solution, as well as the  s+p superconductor solution with both the expectation values for the scalar and vector order parameters non zero. These four kinds of solutions are all possible solutions with one free parameter $T$ at fixed value of $(b,\Delta)$, but the different solutions are in different branches. Thus at some fixed value of $T$, there may exist more than one solution, which means that the system may have more than one possible phase simultaneously. So we need to calculate the free energies of the different phases (solutions) to find out which is the most favored one.

The free energy of the system is equal to the temperature $T$ times the Euclidean on shell action of the bulk spacetime
\begin{equation}
\Omega=T S_{E}.
\end{equation}
In $S_{E}$, we should consider both the bulk terms and some boundary terms. Remember that all the solutions are source free and we have fixed the chemical potential, which means we study the system in grand canonical ensemble. In this case, the action $S_E$ can be expressed as
\begin{eqnarray}
S_{E}&=&-\frac{1}{2\kappa_g^2}\int d^4 x \sqrt{g}\Big[R+\frac{6}{L^2}+2b^2(-\frac{1}{4}F_{\mu\nu}^aF^{\mu\nu a}\nonumber \\ &&
~~~~~~~~~~~~~~~~~~~~ -D_\mu\Psi^a D^\mu\Psi^a-m^2\Psi^a\Psi^a)\Big]  \nonumber \\ &&-\frac{1}{\kappa_g^2}\int_{r \rightarrow\infty} d^3 x \sqrt{h}\Big(K-\frac{2}{L}\Big).
\end{eqnarray}
By substituting the solutions to the Euclidean action, the integrand can be written as a total derivative term, thus the on shell action and the free energy can be evaluated by the boundary value of the functions as

\begin{eqnarray}
\frac{2\kappa_g^2}{V_2} \Omega = \lim\limits_{r\to\infty}
\big[&\frac{2 r^2 N(r) \sigma (r) f'(r)}{f(r)}-r^2 \sigma (r) N'(r)\nonumber\\&
-2 r^2 N(r) \sigma '(r)+4 r^2 \sqrt{N(r)} \sigma (r)\nonumber\\&
-2 r N(r) \sigma (r)\big],
\end{eqnarray}
where $V_2$ denotes the area of the two dimensional transverse space.

For the normal phase which is dual to the AdS-RN black brane solution
\begin{eqnarray}
&\phi(r)=& \mu (1-\frac{r_h}{r})~,\nonumber\\
&~ N(r)=& r^2(1-\frac{r_h^3}{r^3}) + b^2 \frac{\mu^2 r_h^2}{2 r^2}(1-\frac{r}{r_h})~,\nonumber\\
&\sigma(r)=&1~,~f(r)=1.
\end{eqnarray}
 the temperature and free energy are
\begin{equation}
T=\frac{r_h}{4\pi}(3-\frac{b^2 \mu^2}{2 r_h^2})~,~ \frac{2\kappa_g^2}{V_2} \Omega= -r_h^3 - \frac{1}{2} b^2 \mu^2 r_h ,
\end{equation}
while 
for the condensed phases, we have 
\begin{eqnarray}
&T=\frac{r_h}{4\pi}(3\sigma_{h0}-\frac{b^2 \phi_{h1}^2}{2 \sigma_{h0} }- b^2 m^2 \sigma_{h0} \Psi_{3h0}^2)~,\nonumber\\&
\frac{2\kappa_g^2}{V_2} \Omega=-2 M_{b0} .
\end{eqnarray}

Using the above formulas, we can get the free energies of all the solutions and compare these free energies at the same temperature and $(b,\Delta)$. Thus we can draw the condensate of the s-wave and p-wave orders for the most stable phase at any temperature  with fixed $b$ and $\Delta$ (Figures~\ref{cond-b0102}, \ref{cond-b03}, \ref{cond-b04}, \ref{cond-b0506}, \ref{cond-b08}). We can see what kind of phase transitions would occur from these figures. At some values of $b$ and $\Delta$, the condensation behavior is very interesting, we show some typical cases in the next section.

\section{\bf  Condensation and phase transition}\label{sect:PT}
We study the model with eight discrete values of the back reaction strength $b$ from $0.1$ to $0.8$. In each value of $b$, we can fix the value of $\Delta$ to study the phase transitions of the system with a decreasing temperature. We should find all the possible solutions and calculate the relevant free energy to figure out the most stable phase at each temperature. Finally we can plot the condensation behavior  of the s-wave and p-wave orders versus the temperature to show the phase transition at the fixed value of $b$ and $\Delta$.

We can also change the value of $\Delta$ continuously to see what would happen to the stable phases and the phase transitions between them. With the information at different values of $\Delta$, we can extract a phase diagram on the $\Delta-T$ plane at the fixed value of $b$. We will show the eight phase diagrams with different values of $b$ in the next section. In this section we show some interesting and typical phase transition behaviors through the figure of condensation values versus temperature at fixed $b$ and $\Delta$.

\begin{figure*}
\includegraphics[width=8cm] {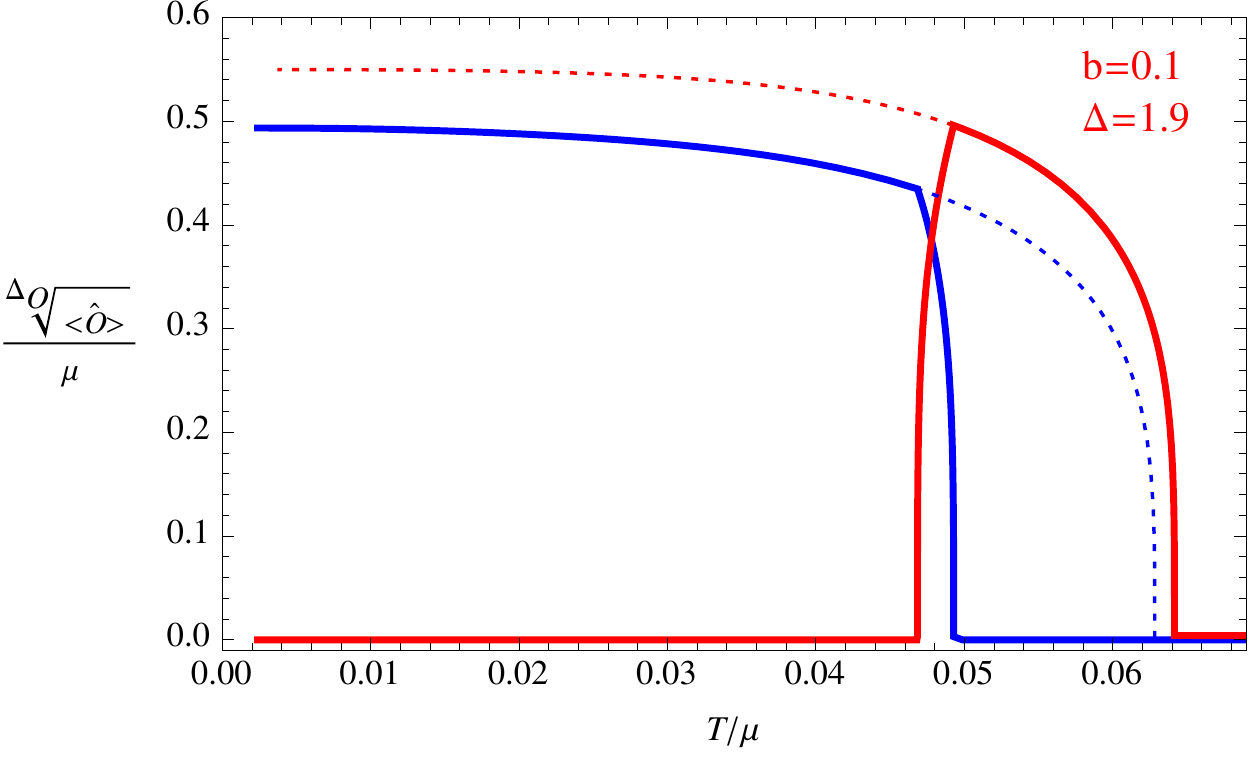}
\includegraphics[width=8cm] {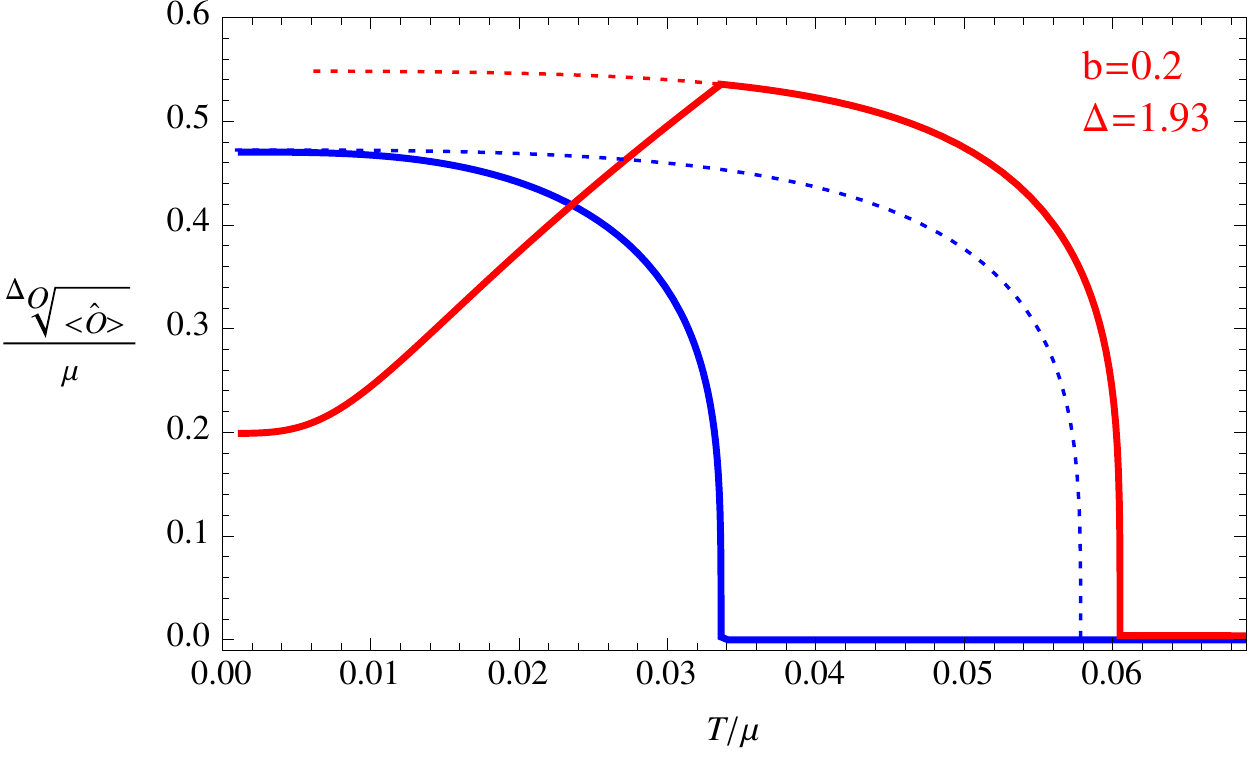}
\caption{\label{cond-b0102}The  condensation value of the s-wave and p-wave orders. The red line denotes the value for the p-wave order, and the blue line denotes the value for the s-wave order.  Dotted lines are for unstable phases and solid lines for stable phases.
}
\end{figure*}

\begin{figure*}
\includegraphics[width=7.8cm] {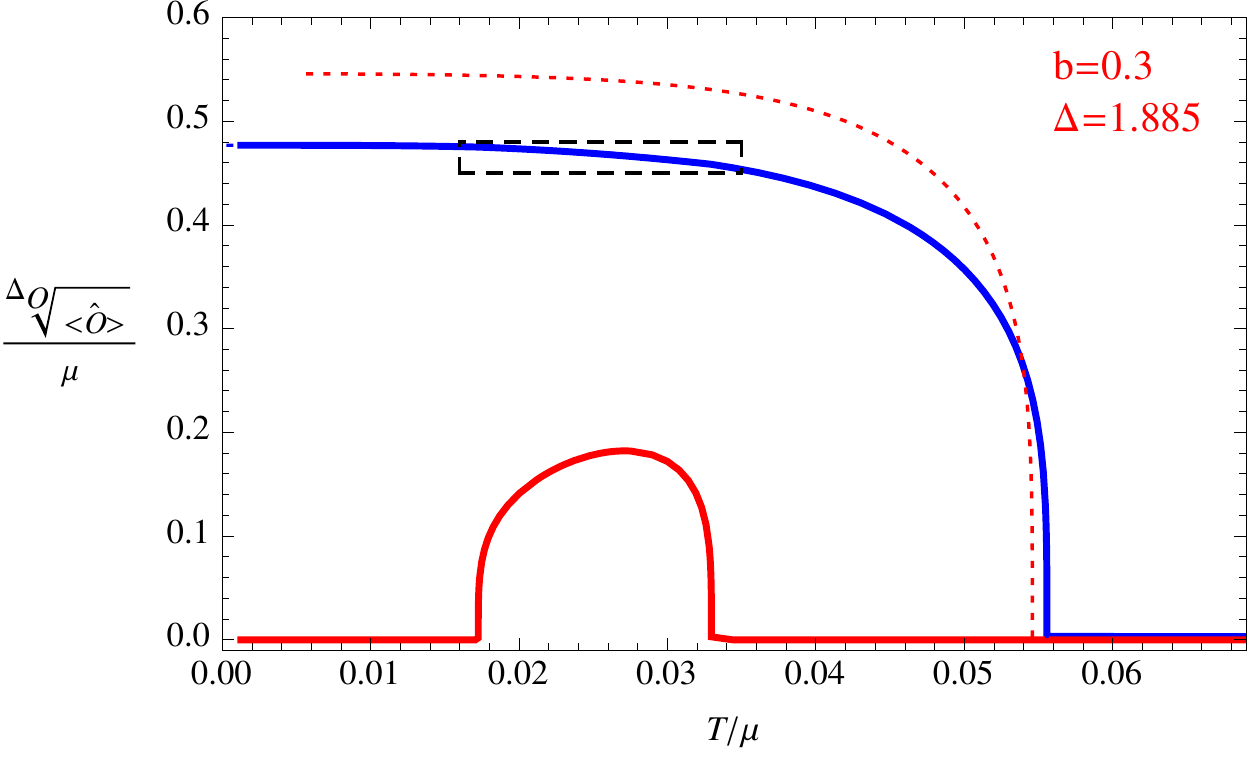}
\includegraphics[width=8.3cm] {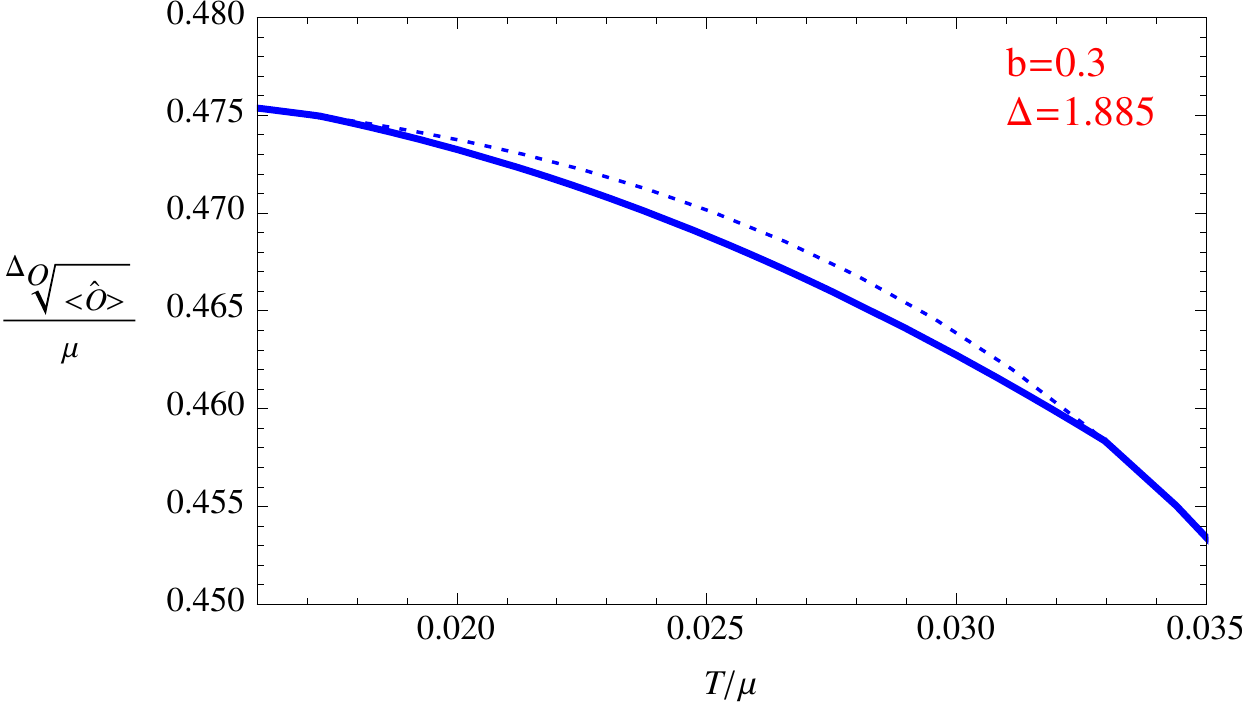}
\caption{\label{cond-b03}The condensation value of the s-wave and p-wave orders. The red line denotes the value for the p-wave order, and the blue line denotes the value for the s-wave order. Dotted lines are for unstable phases and solid lines for stable phases. The right figure is an enlarged version for the rectangle region in the left figure.
}
\end{figure*}

\begin{figure}
\centering
\includegraphics[width=8cm] {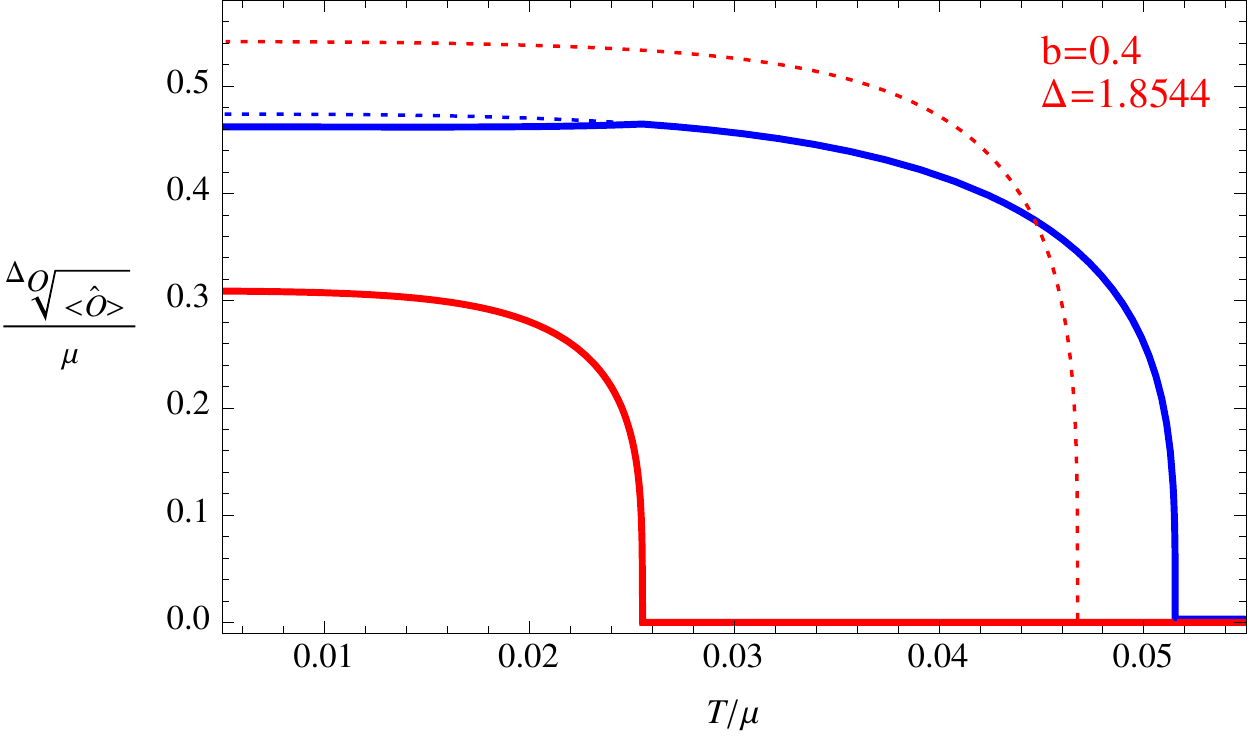}
\caption{\label{cond-b04}The condensation of the s-wave and p-wave orders. The red line denotes the value for the p-wave order, and the blue line denotes the value for the s-wave order. Dotted lines are for unstable phases and solid lines for stable phases.
}
\end{figure}

\begin{figure*}
\includegraphics[width=8cm] {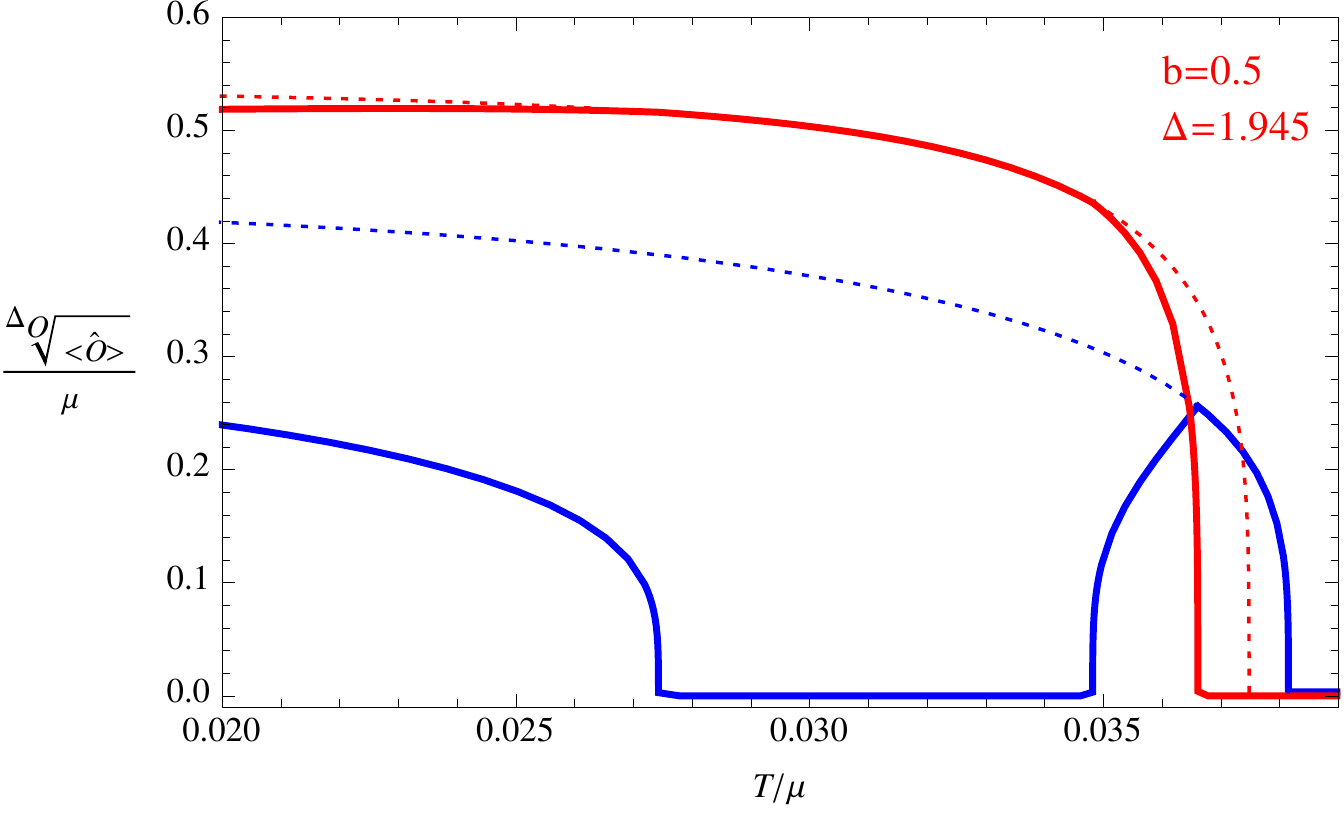}
\includegraphics[width=8cm] {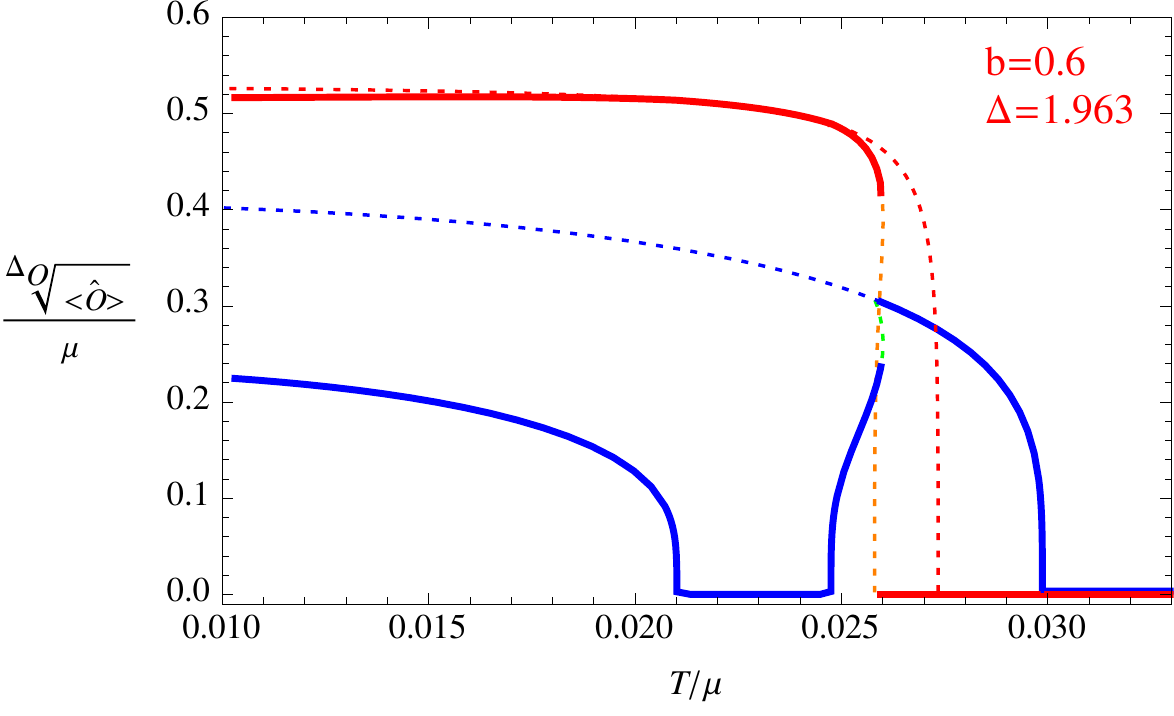}
\caption{\label{cond-b0506}The condensation value of the s-wave and p-wave orders. The red line denotes the value for the p-wave order, and the blue line denotes the value for the s-wave order. Dotted lines are for unstable phases and solid lines for stable phases. The dotted orange line and dotted green line in the right figure denote the condensation value for the p-wave and s-wave orders in the unstable s+p solution.
}
\end{figure*}

\begin{figure}
\includegraphics[width=8cm] {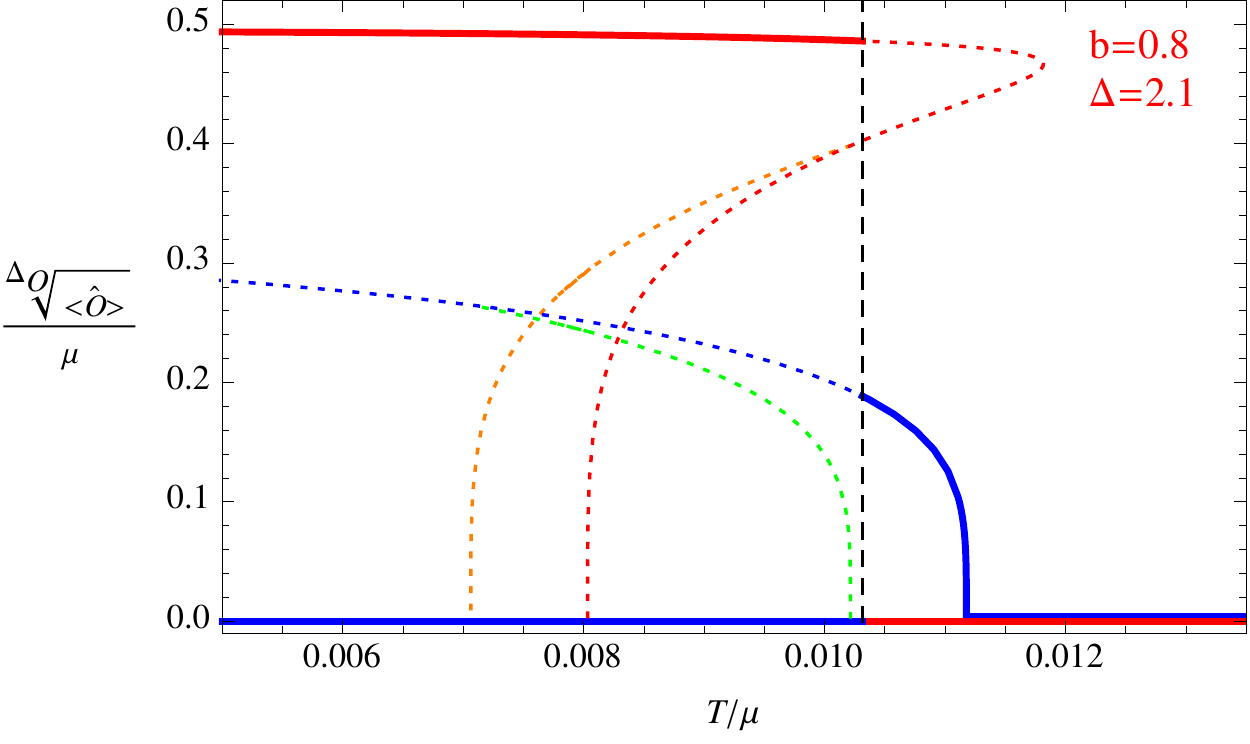} 
\includegraphics[width=8cm] {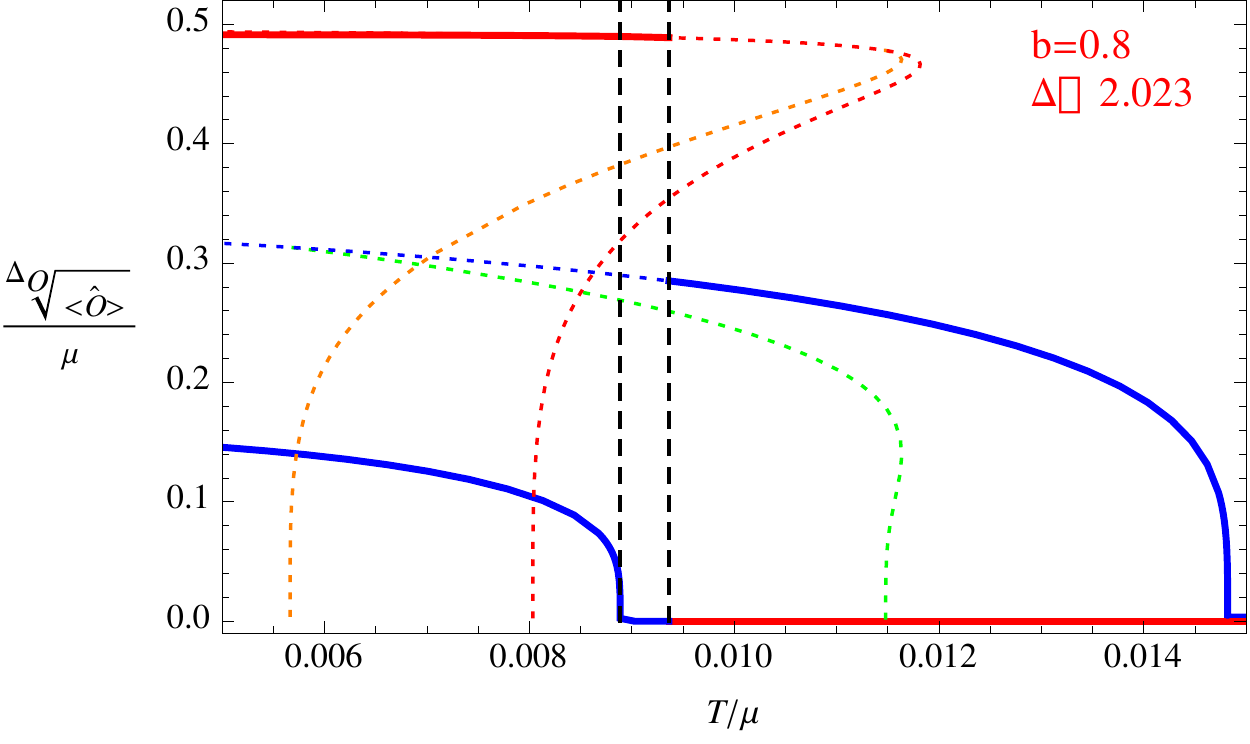}
\includegraphics[width=8cm] {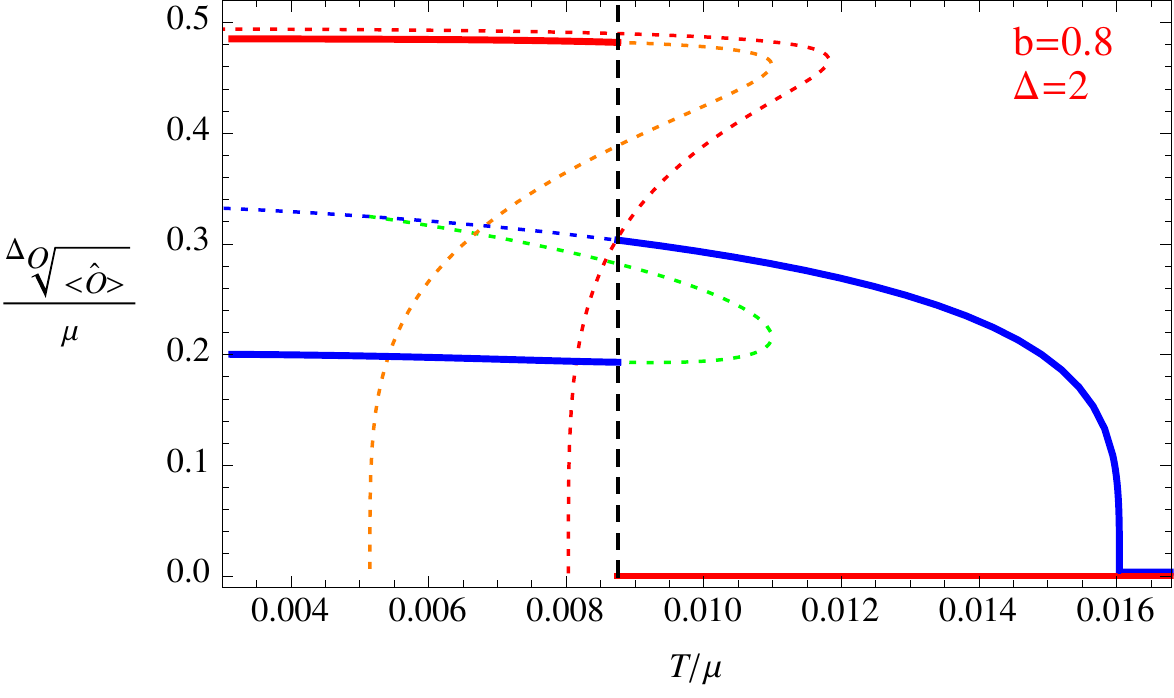}
\caption{\label{cond-b08}The condensation value of the s-wave and p-wave orders. The red line denotes the value for the p-wave order, and the blue line denotes the value for the s-wave order. Dotted lines are for unstable phases and solid lines for stable phases. The dotted orange line and dotted green line denote the condensation value for the p-wave and s-wave orders in the unstable part of the s+p solution. We use the vertical dashed black lines to mark the critical temperatures between the condensed phases clearly.
}
\end{figure}

For each value of $b$, we vary  the value of $\Delta$ to find all  possible phase transitions. Below we pick some typical values of $b$ and $\Delta$, and draw figures for the condensation values of the s-wave and p-wave operators versus temperature in Figures~\ref{cond-b0102}, \ref{cond-b03}, \ref{cond-b04}, \ref{cond-b0506}, \ref{cond-b08}. In these figures, we use the solid red line to show the condensation value of the p-wave operator, and  the solid blue line to show the condensation value of the s-wave operator. We also use dotted lines to show the condensation values in unstable branches. The dotted red lines denote the condensation value of the p-wave order in the unstable section of the p-wave phases, and the dotted blue lines denote the value of the s-wave order in the unstable section of the s-wave phases. The s+p coexisting phase might also become unstable in our study. So in the unstable s+p phases, we use dotted orange line to denote the condensation value of the p-wave order and use the dotted green line to denote that of the s-wave order. In Figure~\ref{cond-b08} we also use vertical dashed black lines to denote critical temperatures of phase transitions between different condensed phases.

In Figure~\ref{cond-b0102}, the left plot is for the case $b=0.1$ and $\Delta=1.9$. In this figure, we can see a typical behavior of the s+p system. In this case the s+p coexisting phase connects the p-wave phase in higher temperatures and the s-wave phase in lower temperatures. This behavior is similar to that in the probe limit~\cite{Nie:2013sda}, but here the temperature region of the s+p coexisting phase is larger than that in the probe limit.

The right plot of Figure~\ref{cond-b0102} shows the case with $b=0.2$ and $\Delta=1.93$. We can see that the s+p coexisting phase emerges from a p-wave phase and is stable in the low temperature region (the condensate of the p-wave order does not vanish even in low temperature). In the s+p coexisting phase, the condensation value of p-wave order is reduced when the value of the s-wave order increases from zero. This means that the s-wave order and the p-wave order  still repel each other.

Figure~\ref{cond-b03} shows the results with $b=0.3$ and $\Delta=1.885$. The right plot is an enlarged version of the rectangle region in the left one. In this case, we find an interesting behavior for the s+p coexisting phase. The s+p coexisting phase begins from an s-wave phase at a higher temperature and finally goes back to the s-wave phase at a lower temperature. We can see from the picture that the condensation value of the p-wave order forms a shape of the letter ``n'', so we call this as an ``n-type" condensation. In the n-type phase transition, the order parameter of the p-wave only gets non-zero values in a small region. Note that we can see from the right plot that the value of s-wave order in the s+p coexisting phase is reduced, compared to the case of the pure s-wave phase.

In Figure~\ref{cond-b04}, we show the results with $b=0.4$ and $\Delta=1.8544$. In this figure, we can see that similar to the case in the right plot of Figure~\ref{cond-b0102},  the s+p coexisting phase is also stable in the low temperature region, but at this time, the s+p coexisting phase emerges from an s-wave phase instead.

Figure~\ref{cond-b0506} shows two similar cases. The left plot is for $b=0.5,~\Delta=1.945$ and the right one for $b=0.6,~\Delta=1.963$. There are five regions of different phases in each of the plot, from the right to the left, they are the normal phase, the s-wave phase, the s+p coexisting phase, the p-wave phase and the second branch of s+p coexisting phase. The main difference between the two cases is that in the left plot, the phase transition from the s-wave phase to the s+p coexisting phase is second order, but in the right plot, this phase transition is first order. We can see from the two plots that the system goes into the p-wave phase from the s+p coexisting phase and goes back to the s+p coexisting phase at a lower temperature.  The shape of the solid blue line forms a letter ``u", so we call this as a ``u-type" condensation. These kinds of phase transitions are known as ``reentrant" phase transitions in condensed matter physics, and have also been found in holographic study in Ref.~\cite{DG}.

Figure~\ref{cond-b08} is for the cases with $b=0.8$ and three different values of $\Delta$. In these plots, we use the vertical dashed black lines to denote the critical temperature of the phase transitions between different condensed phases. The first plot shows the case $\Delta=2.1$. We can see that in this case, there is no stable s+p phase and a first order phase transition occurs between the s-wave phase and the p-wave phase. However, an unstable s+p solution  exists, and the condensation values of s-wave and p-wave orders in the unstable s+p phase are denoted by the dashed green and dashed orange lines.

The second plot of Figure~\ref{cond-b08} presents the case with $\Delta=2.023$. In this figure, we can see that there is also a first order phase transition between the s-wave phase and the p-wave phase. And similar to the case with $\Delta=2.1$, there is an unstable s+p solution connecting the s-wave phase and the p-wave phase. Besides this unstable s+p solution, there is another branch of stable s+p phase in the low temperature region. It is also interesting to notice that the two s+p phases overlap in some temperature region, which means one might find two different branches of solutions for the s+p coexisting phase at some  temperatures. We should also notice that the condensation value of the p-wave order in the lower temperature branch of s+p phase differs from the value in the p-wave phase, although the two red lines (solid one for the s+p phase and dotted one for the p-wave phase) are very close to each other.

The third plot in Figure~\ref{cond-b08} shows the case with $\Delta=2$. We can see that the p-wave phase in this case is always unstable while the s+p phase is partly unstable, and the phase transition from the s-wave phase to the s+p phase is first order.

From the above figures we can see that this holographic system exhibits  a rich phase structure. These different phase transition behaviors can be used to look for and test universal behaviors in holographic systems. In particular, we have found that in this model, with some choice of the value of parameters, an s+p solution might be unstable, which is quite different from the previous studies, where once an s+p solution appears, it is usually stable.  In the next section we will construct the phase diagram of the system in terms of the dimension of the scalar order and the temperature. Before that, in the rest of this section, we will discuss the potential role of the unstable s+p solution in phase transitions between the s-wave and p-wave phases.

\subsection*{The Swallow Tail for First Order Phase Transition}

Note  that to get the above condensation behaviors, we have calculated the free energies for all the phases to confirm the stability in each case. The critical temperatures for the first order phase transitions are also calculated from the free energy of the two related phases. From the free energy curves, we found that an s+p solution always exists when the two free energy curves for the s-wave phase and p-wave phase intersects. But this s+p solution is not always stable. We classify the three cases for the stability of this s+p solution as follows:
\begin{enumerate}
\item $Stable$: \label{stableSP} In this case, there exist two second order phase transitions to and from the s+p phase, instead of a first order phase transition between the s-wave and p-wave phases.
\item $Totally~Unstable$:\label{UnstableSP} A first order phase transition between the p-wave phase and the s-wave phase occurs.
\item $Partly~Stable~ \&~ Partly~ Unstalbe$: \label{partlySP} A first order phase transition occurs between the s+p phase and the s-wave or p-wave phase.
\end{enumerate}

In our previous work~\cite{Nie:2013sda}, we have shown the free energy curves for Case \ref{stableSP}, where this s+p solution is stable. It is natural that in this case the s+p solution exists and is the most stable phase in its temperature region. But in the last two cases, especially in the second case, the s+p solution becomes totally unstable. Then we may ask what is the meaning for the existence of the unstable s+p solution?

\begin{figure*}
\includegraphics[width=8cm] {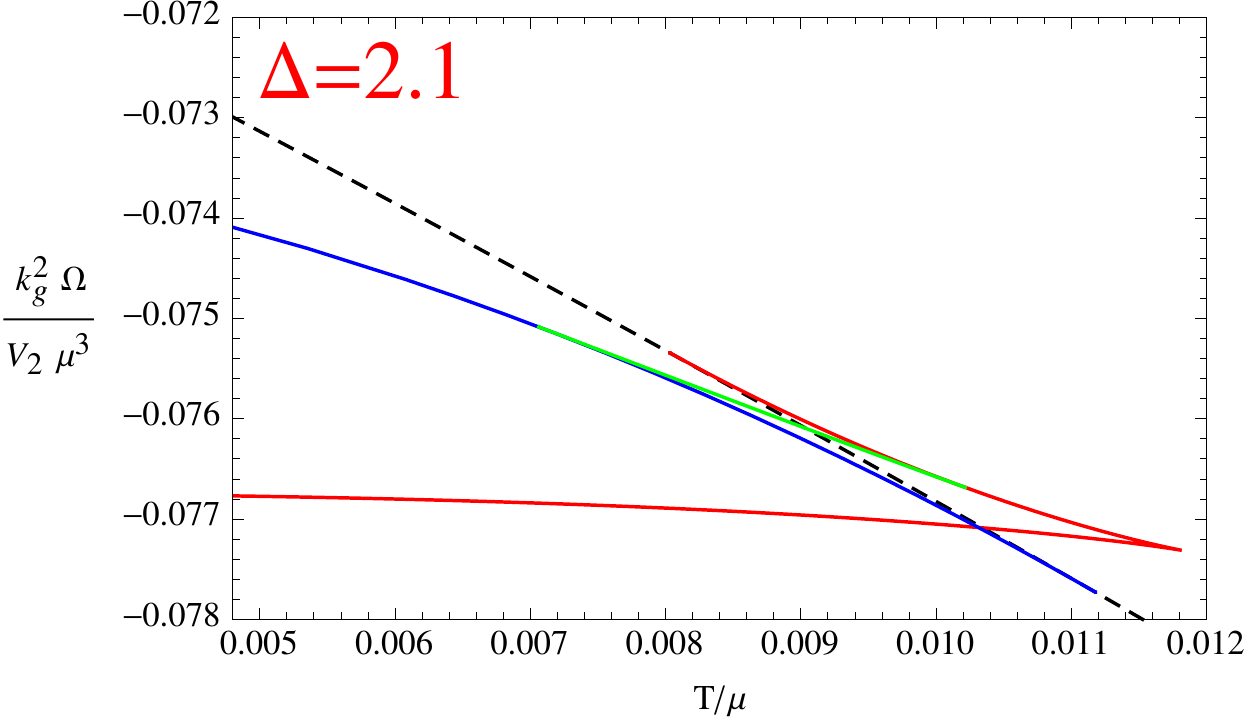}
\includegraphics[width=8cm] {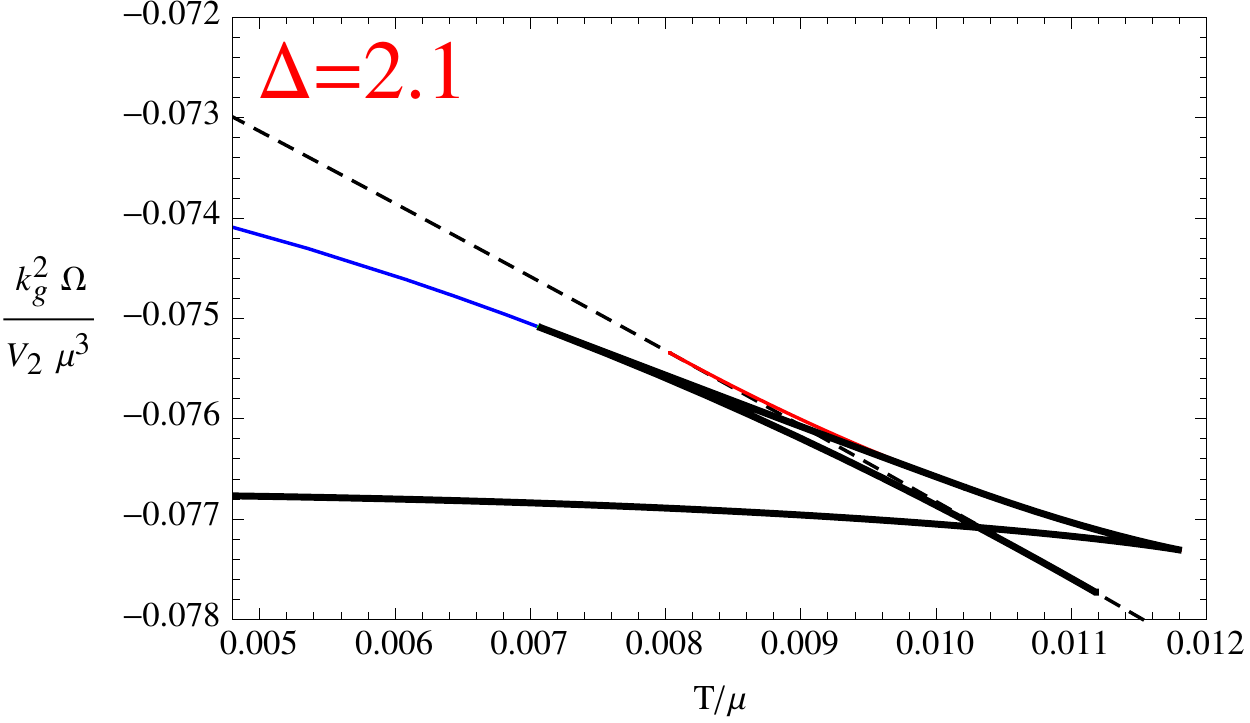}\\
\includegraphics[width=8cm] {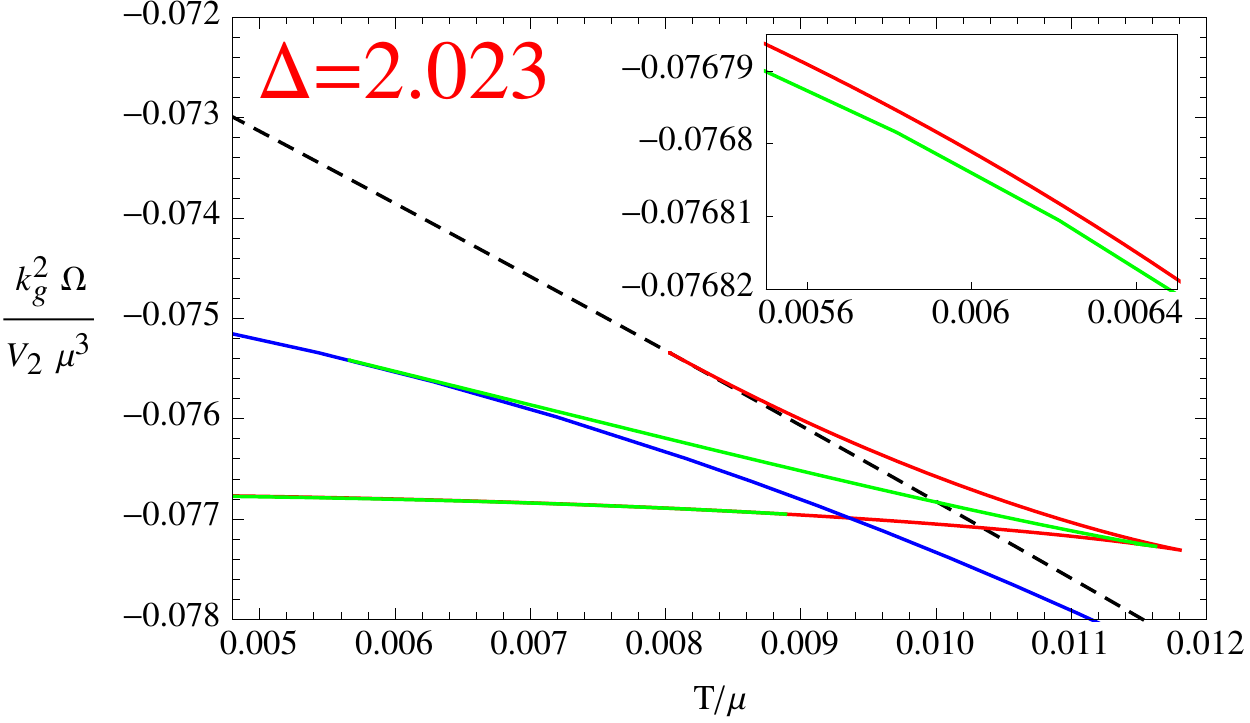}
\includegraphics[width=8cm] {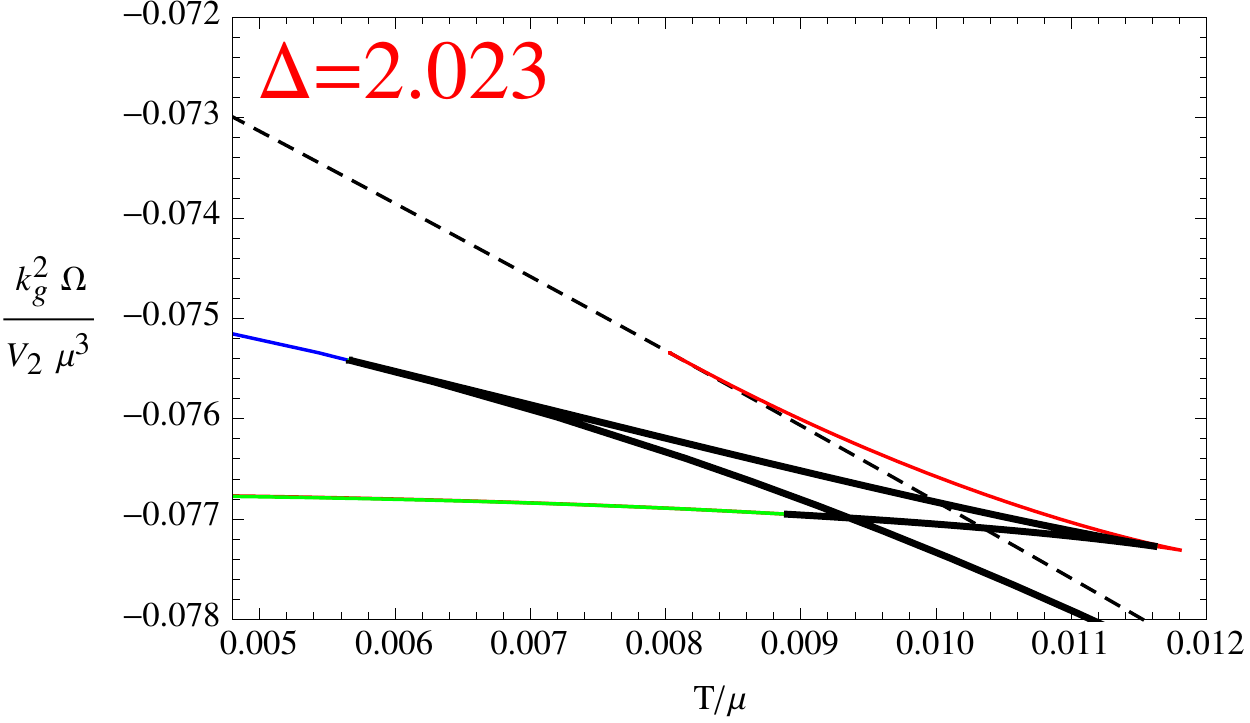}\\
\includegraphics[width=8cm] {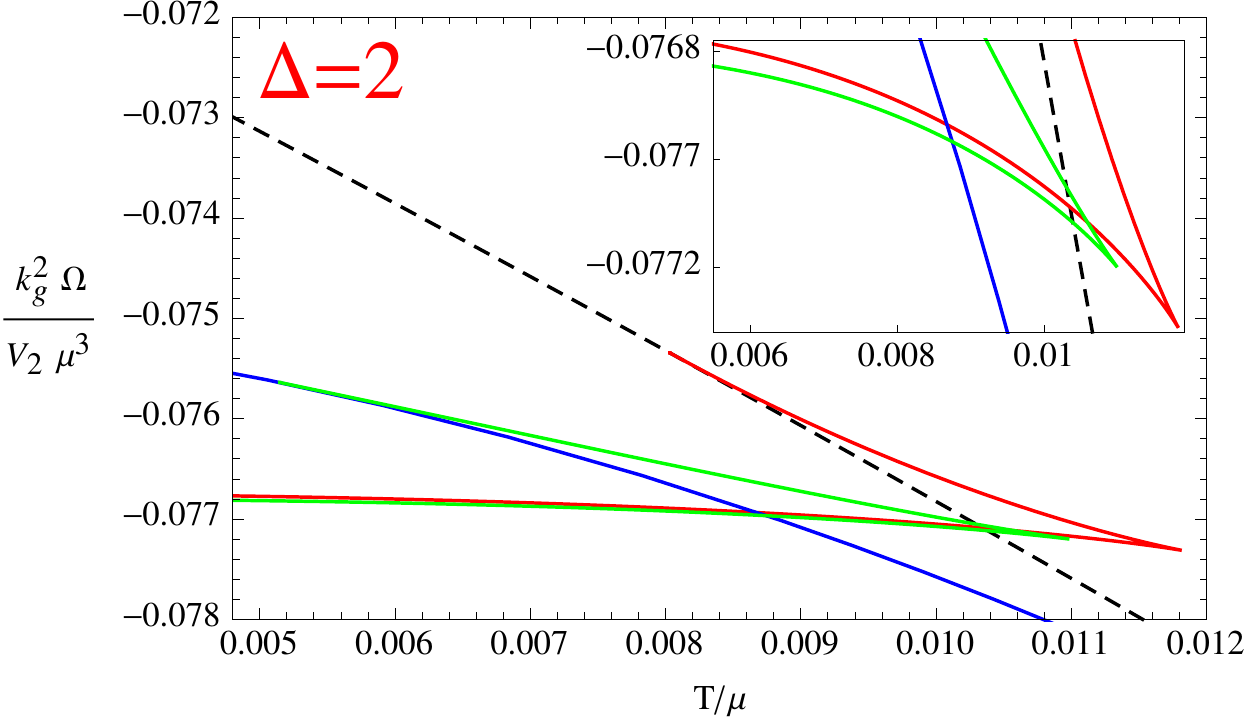}
\includegraphics[width=8cm] {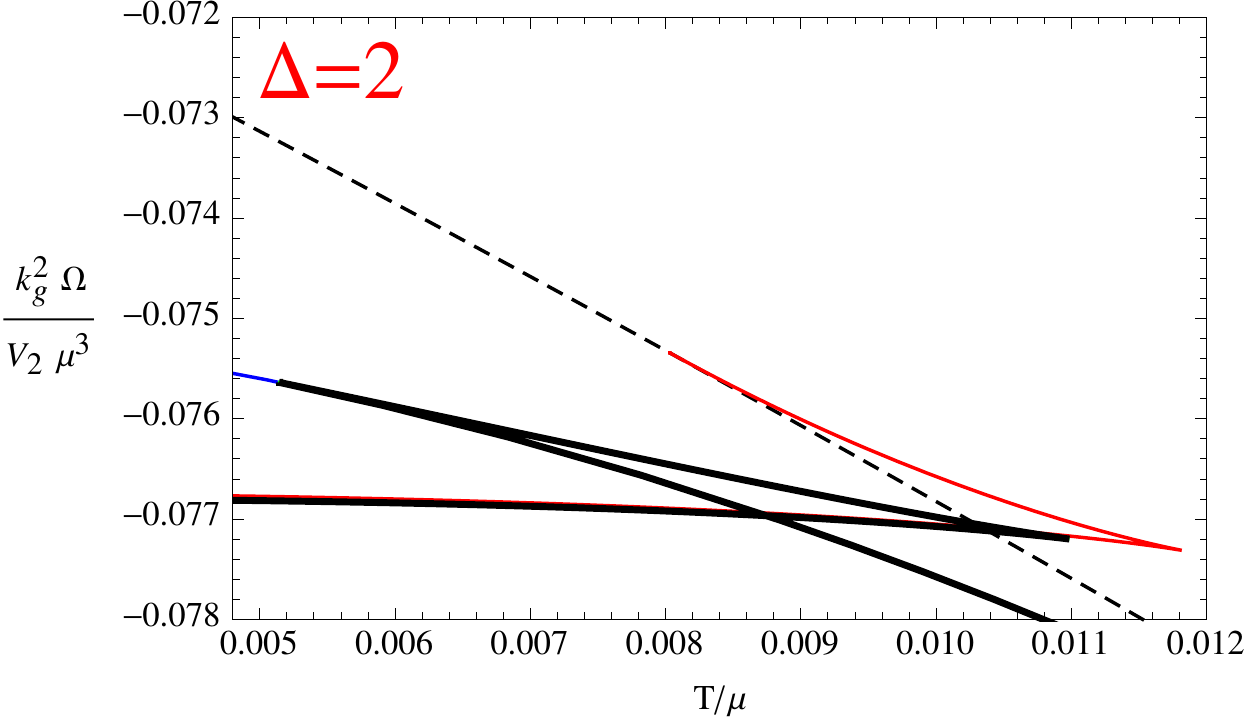}
\caption{\label{FreeE}The free energy of the different phases with $b=0.8$. The first two figures are for the case $\Delta=2.1$, the third and forth figures are for the case $\Delta=2.023$ and the fifth and sixth ones for $\Delta=2$. In the three figures on the left side, the dashed black line denotes the free energy for the normal phase, the solid red line and solid blue line denotes the free energy of the p-wave phase and s-wave phase respectively, the solid green line denotes the free energy of the s+p phase(solution). We also use enlarged version in the third and fifth figure to show the detailed relation between the red and green lines when they are close to each other. In the three figures on the right side, we use thick black curves to show how the free energy curve of different phases form the shape of a swallow tail.}
\end{figure*}

To answer the question, we draw the free energy curves  in Figure~\ref{FreeE} for the three condensation behaviors shown in Figure~\ref{cond-b08}. The two plots in the first line are for the case $b=0.8, \Delta=2.1$, the two plots in the second line are for the case $b=0.8, \Delta=2.023$, and the last two figures are for the case $b=0.8, \Delta=2$. In the plots on the left, we use the solid red line and solid blue line to denote the free energy of the p-wave phase and s-wave phase respectively. The green line denotes the free energy for the s+p phase(solution) and the dashed black line denotes that for the normal phase. We also draw the enlarged version for the last two cases to see which one is lower when the green and red lines are very close to each other in some region. We use the same data to draw the plots on the right side and those on the left side, and we use thick black curves to denote the swallow tail shape in first order phase transitions in the plots on the right side.

We can see that when the s+p solution near the free energy intersection point becomes partly or totally unstable, a first order phase transition occurs. This first order phase transition can be from the s-wave phase to the p-wave phase as in the case when the s+p solution is totally unstable, or from the s-wave phase to the s+p phase as in the case when the s+p solution is partly unstable. In both the two cases, we can find a classical swallow tail in the free energy curves for the first order phase transition. To make this clear, we use thick black curves to mark the swallow tail shape in the plots on the right side of Figure~\ref{FreeE}. Remember that we use the same data to draw the left and right figures on the same row. One can compare the left and right plots to see that although the unstable part of the s+p solution is not experienced in the phase transition, it is still quite necessary in forming the swallow tail shape of the free energy curve. It seems that the existence of the unstable s+p solution is to bridge the s-wave phase to the p-wave phase and to form the standard swallow tail shape of free energy for the first order phase transitions.

According to the previous results, the swallow tail shaped free energy curve seems always existing in any first order phase transition. If we assume the swallow tail shape of free energy curve as a universal property for the first order phase transition, we can explain the existence of s+p solution near the free energy intersection point as follows. When the free energy curves of the s-wave phase and the p-wave phase have an intersection point as in Figure~\ref{FreeE}, the s+p solution must exist. Otherwise a first order phase transition would occur between the s-wave phase and p-wave phase, and without the s+p solution, the free energy curve can not form a swallow tail shape, which contradicts with the universality we assumed. This contradiction could be solved by an s+p solution near the intersection point of free energy curves. This s+p solution could either be (partly) stable to avoid the first order phase transition between the s-wave phase and p-wave phase, or be totally unstable but bridge the s-wave and p-wave phases and form the swallow tail shape of free energy curve for the first order phase transition between the s-wave and p-wave phases. We can also use the same logic to explain the existence of the s+d phase in the holographic s+d system studied in Ref.~\cite{Nishida:2014lta}. If our assumption for the swallow tail shape in first order phase transitions is true, it can be very useful in finding new solutions near some free energy intersection points in more general systems.

\section{Phase diagrams}\label{sect:PD}
In the previous section, we have shown the condensation behaviors for some interesting cases. If we get all the condensation behaviors for different values of $\Delta$ at fixed $b$, we can summarize the information at different $\Delta$ to build a $\Delta-T$ phase diagram at this fixed value of $b$. In this section, we use the above method to give the $\Delta-T$ phase diagrams with eight different values of the back reaction parameter $b$ from $0.1$ to $0.8$. With these phase diagrams, we can understand the rich phenomena  shown in the previous section much more clearly. And we can also compare the phase diagrams to study the effect of back reaction concretely.

\begin{figure*}
\includegraphics[width=8cm] {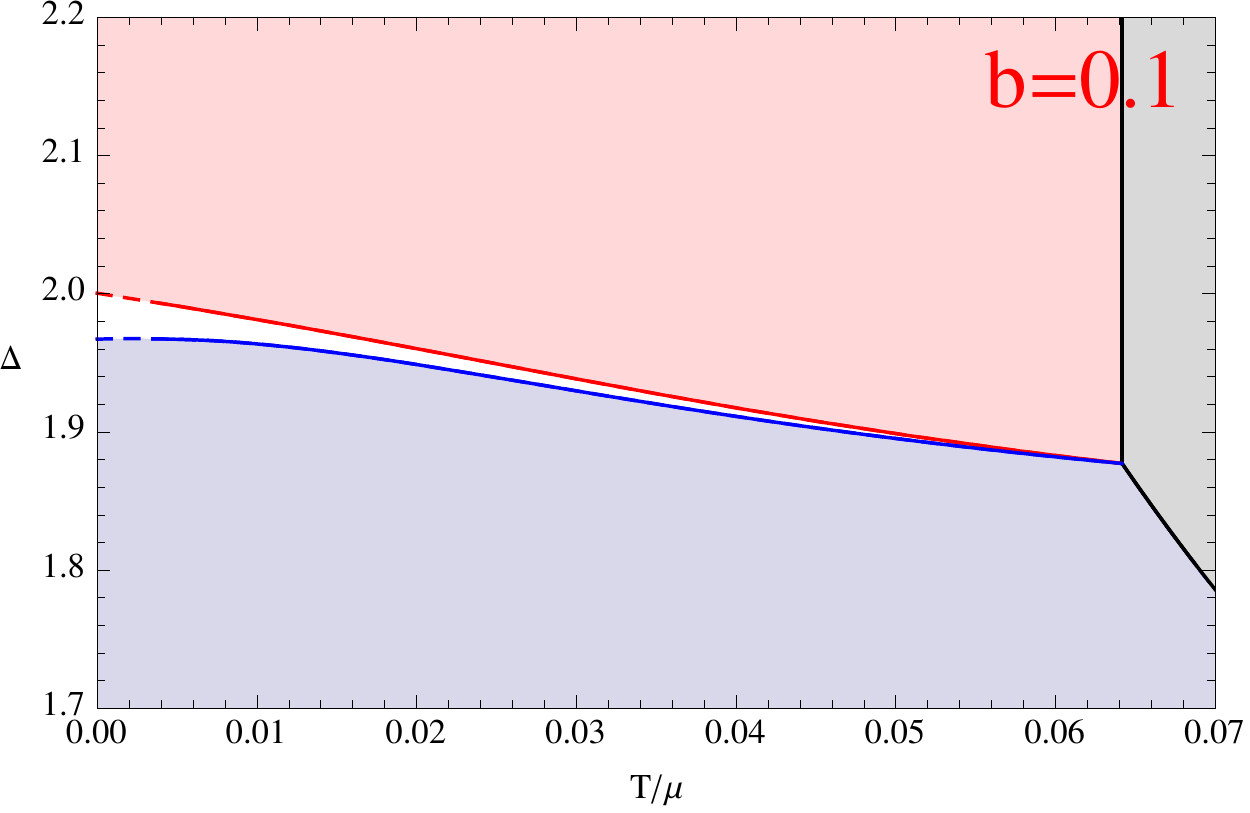}
\includegraphics[width=8cm] {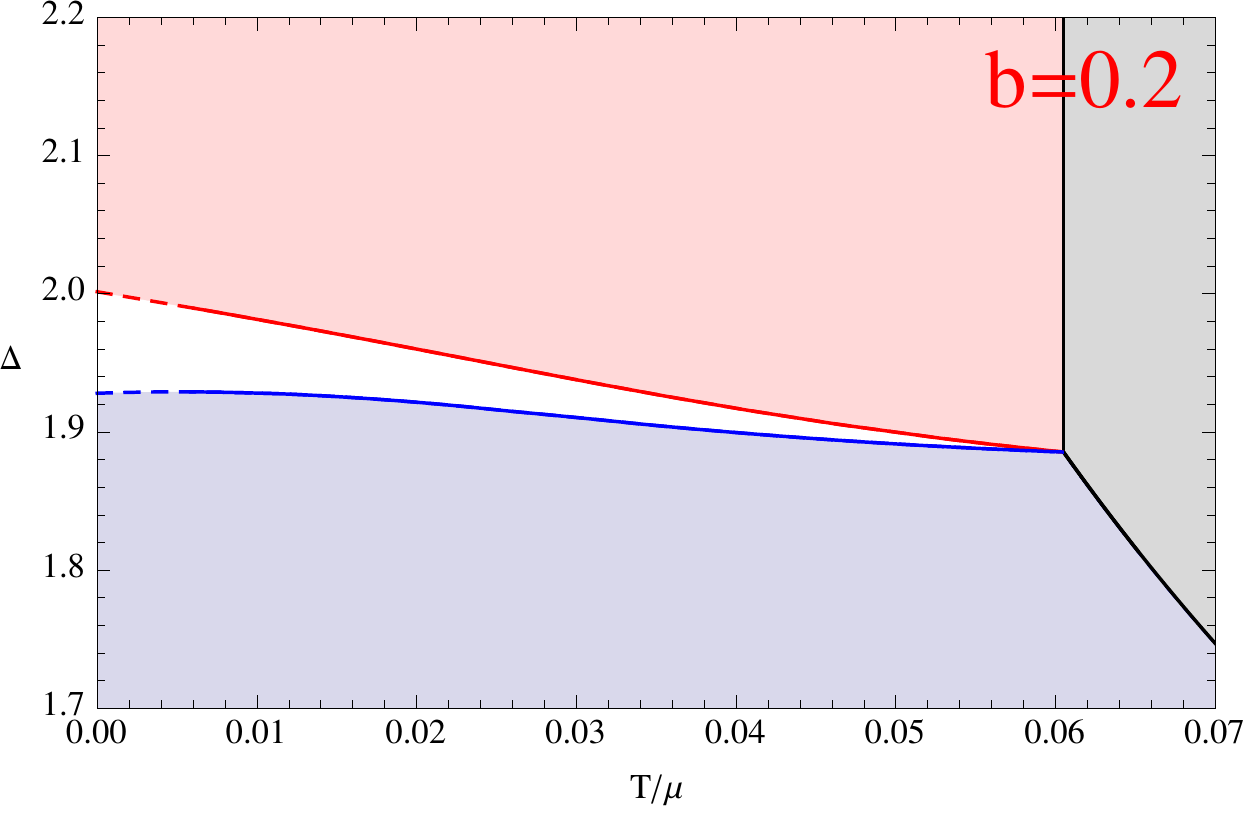}\\
\includegraphics[width=8cm] {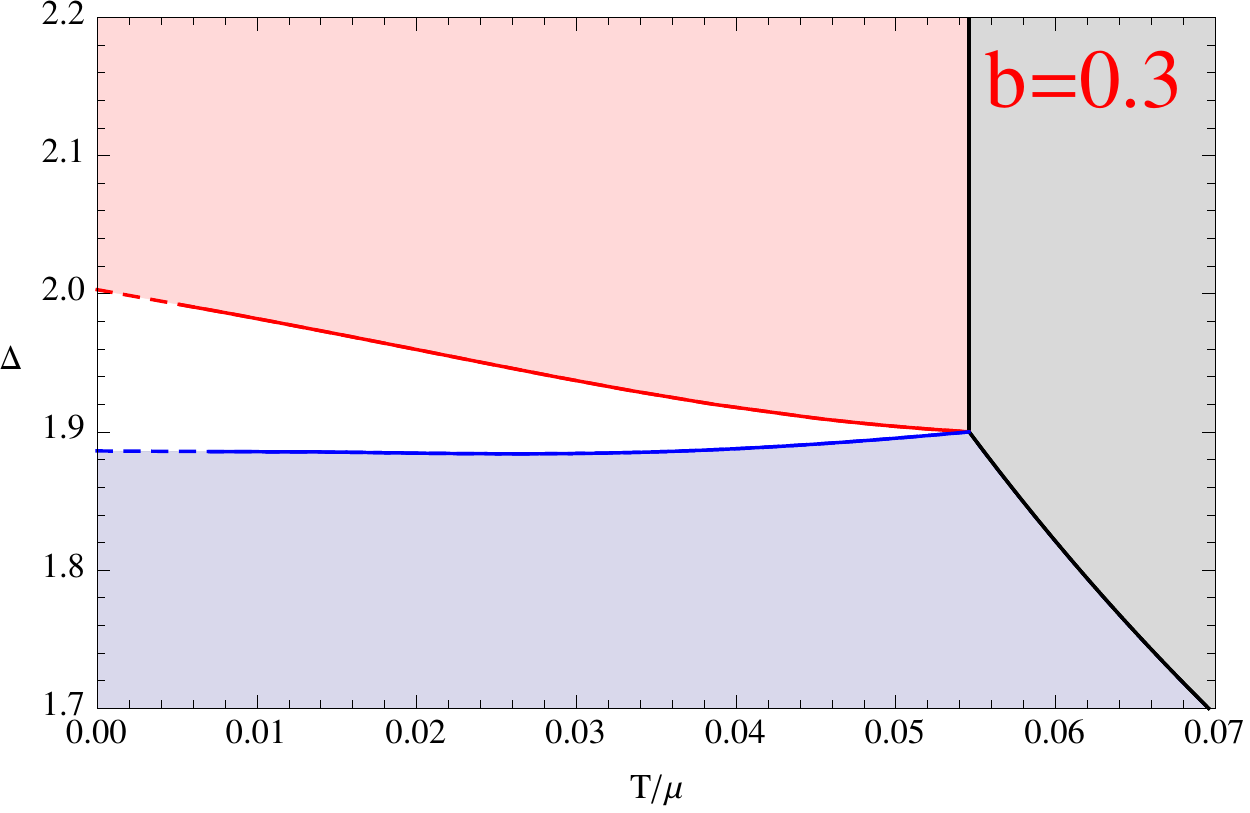}
\includegraphics[width=8cm] {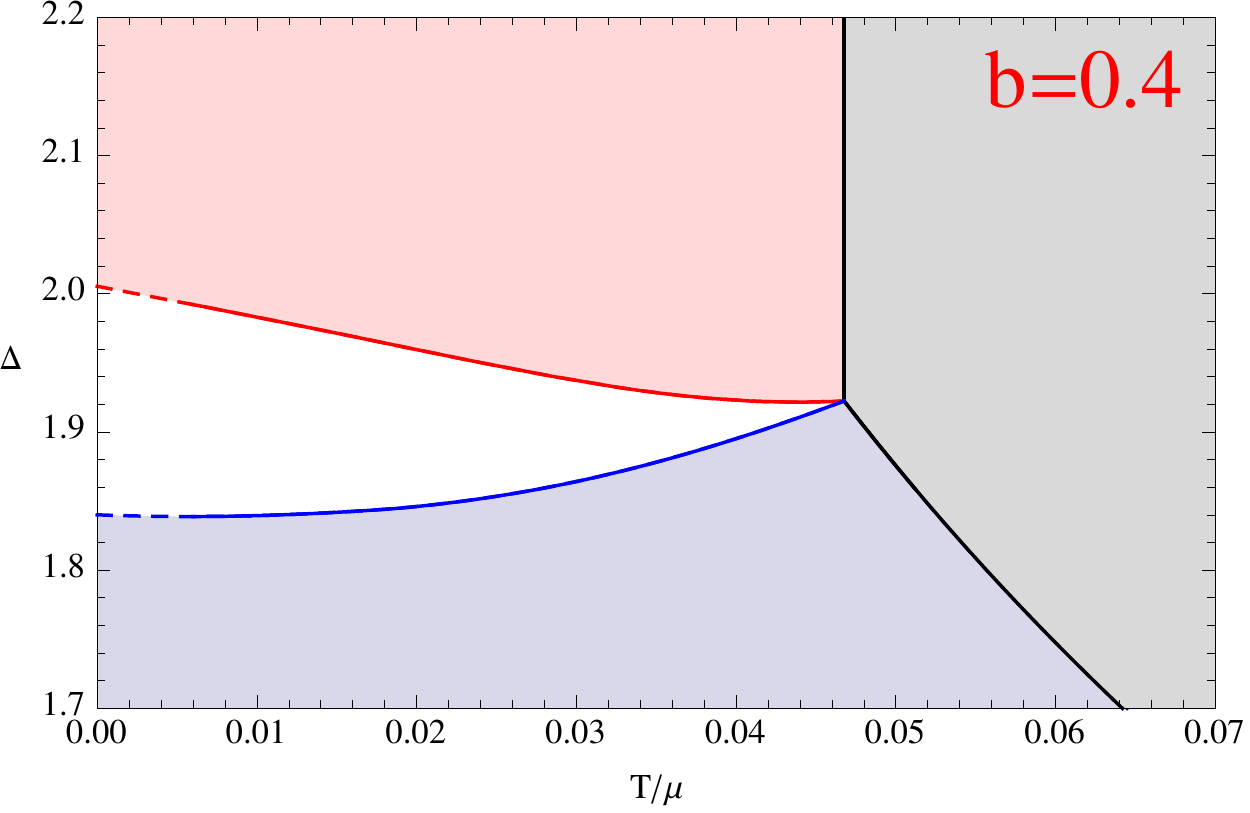}
\caption{\label{phasediagram1}The $\Delta-T$ phase diagrams for $b$ values from $0.1$ to $0.4$. In these figures, the areas colored by light gray, light blue, light red and white denote the normal phase, the s-wave phase, the p-wave phase and the s+p coexisting phase respectively. The black curves are made up by phase transition critical points between the s-wave or p-wave phases and the normal phase. The solid red curves describe the second order phase transition critical points between the p-wave phases and the s+p coexisting phases. The solid blue curves represent the second order phase transitions between the s+p coexisting phases and the s-wave phases. The dashed lines near the zero temperature regions are extrapolated from data in higher temperature region.}
\end{figure*}

\begin{figure*}
\includegraphics[width=8cm] {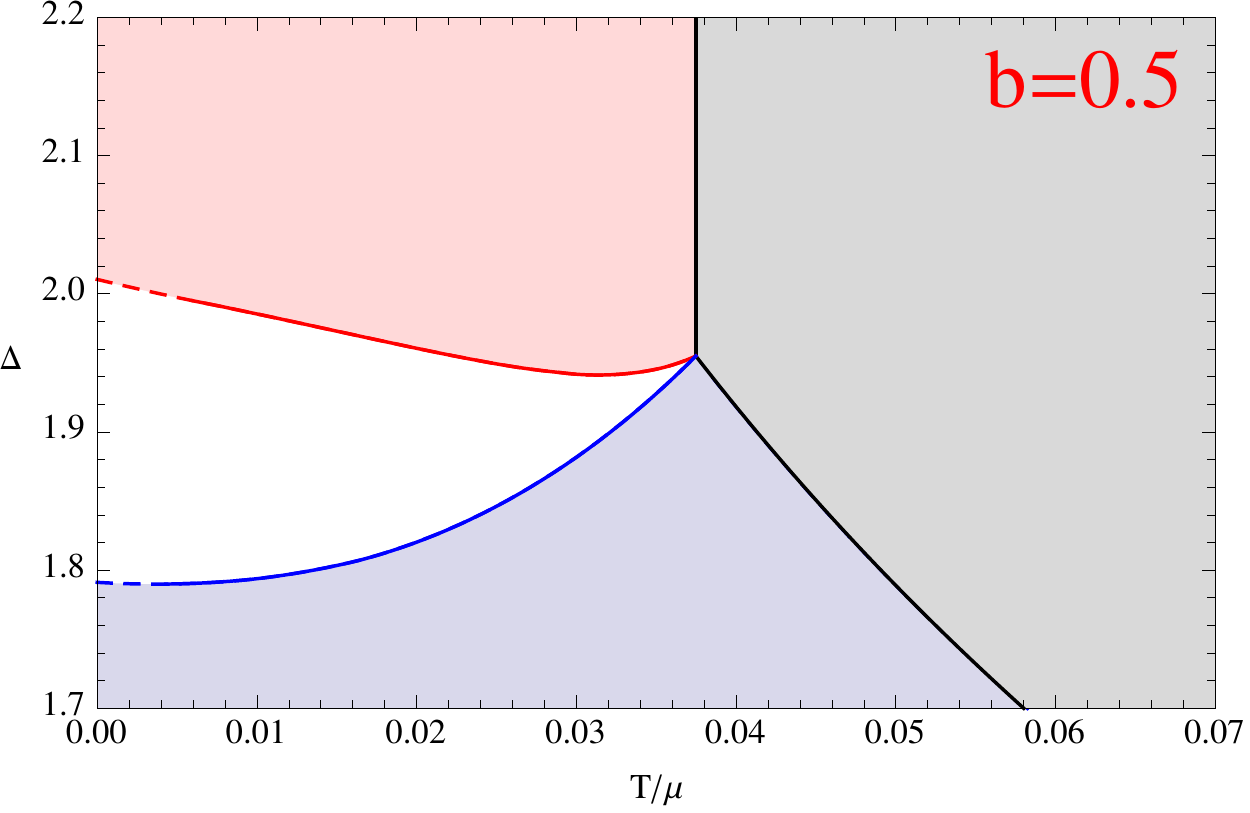}
\includegraphics[width=8cm] {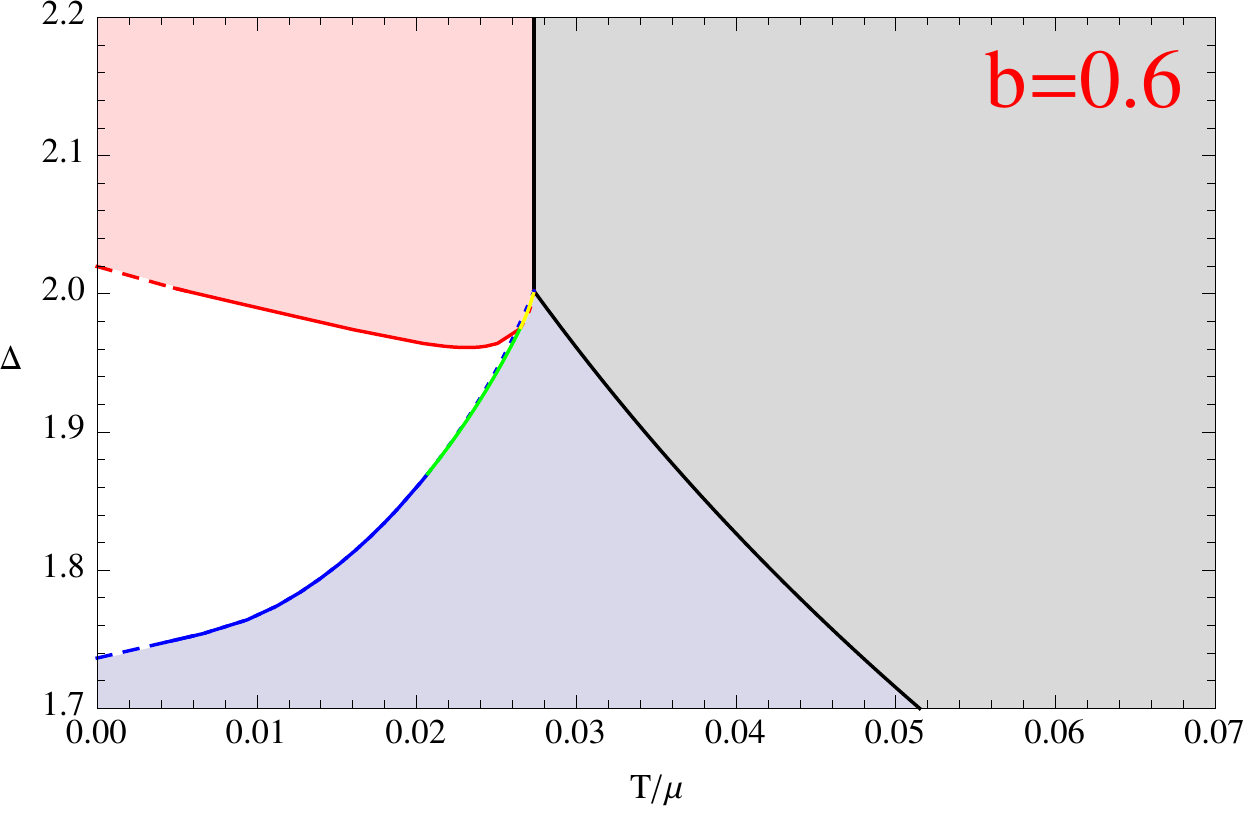}\\
\includegraphics[width=8cm] {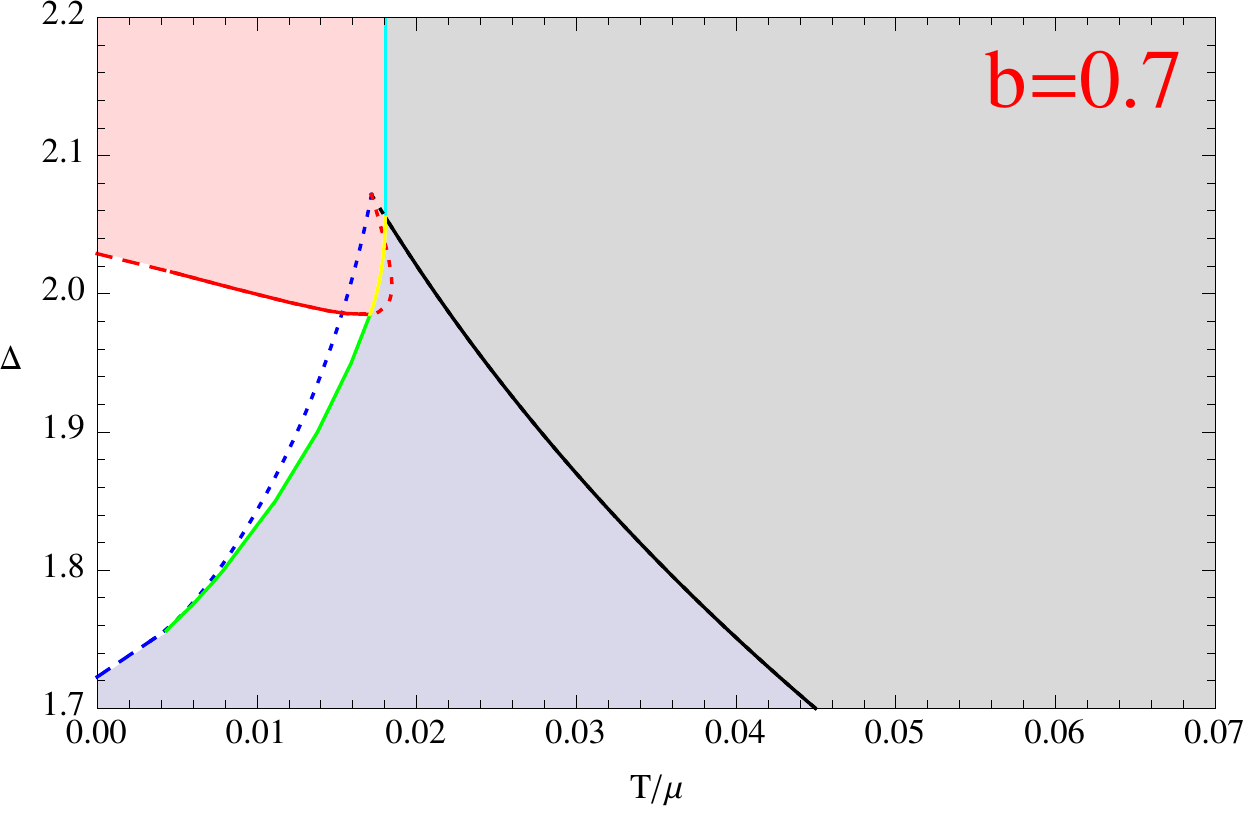}
\includegraphics[width=8cm] {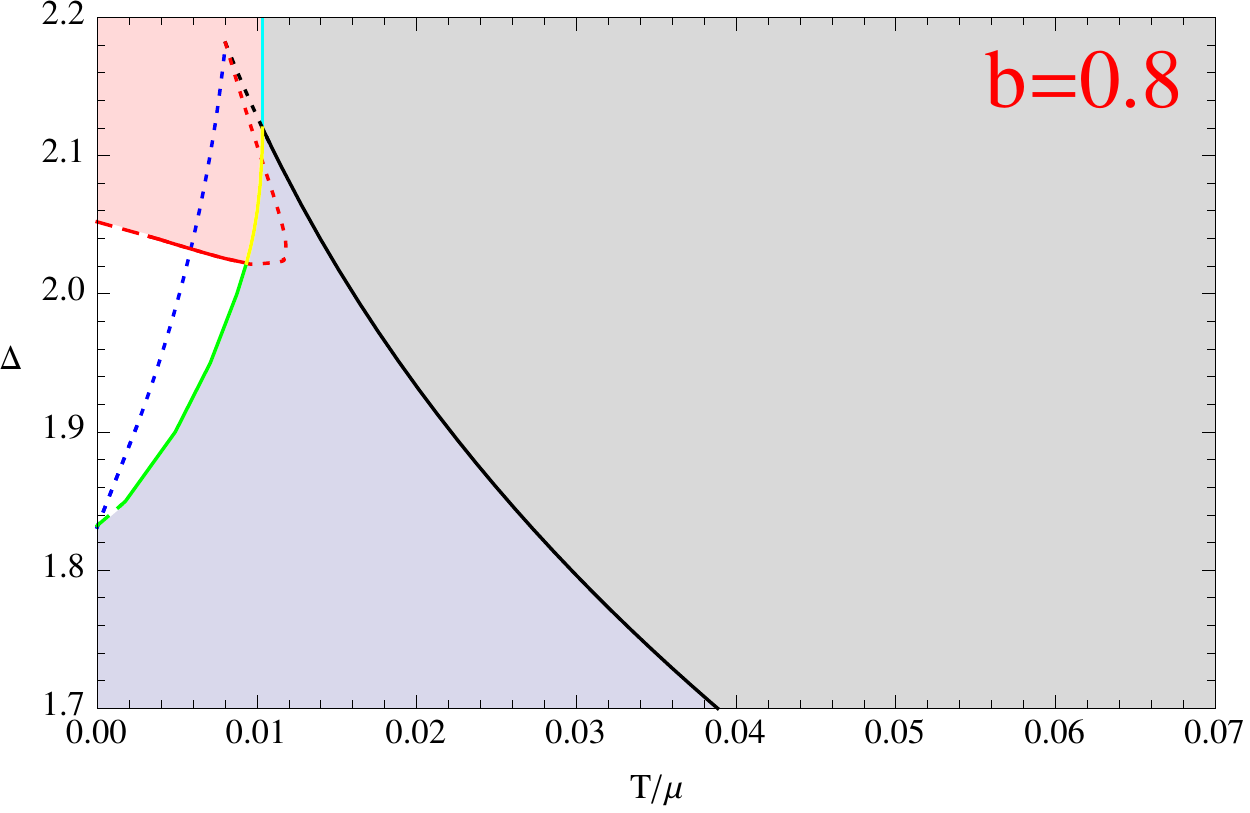}
\caption{\label{phasediagram2}The $\Delta-T$ phase diagrams for $b$ values from $0.5$ to $0.8$. In these figures, the solid cyan line denotes the critical points of the first order phase transitions from the normal phase to the p-wave phase. The solid yellow lines represent the first order phase transition critical points from s-wave phases to p-wave phases. The solid green lines represents the first order phase transition critical points from s-wave phases to s+p coexisting phases. The dotted part of the red and blue lines are the emergent points of the unstable part of s+p phases. Notations of other lines are the same as introduced in Figure~\ref{phasediagram1}.}
\end{figure*}

We show all the phase diagrams in Figure~\ref{phasediagram1} and Figure~\ref{phasediagram2}, respectively.  In the eight phase diagrams with $b$ from $0.1$ to $0.8$, we can see that all the diagrams have four different phases. These four phases are the normal phase, the s-wave phase, the p-wave phase and the s+p coexisting phase colored by light gray, light blue, light red and white, respectively. The thermodynamic stability of all these phases has been confirmed by our calculation of free energy.

We also use different lines in the figures to show the sets of different phase transition critical points or emergent points of unstable phases. The solid lines divide the different phases, and the dashed lines near the zero temperature region are obtained  by extrapolation from the solid lines with the same color. The dotted lines denote the points where an unstable solution emerges. We use different colors to mark the different critical lines. The second order s-wave and p-wave phase transitions from the normal phase have been well studied, we color the critical lines black. When the back-reaction is strong enough, the p-wave phase transition  becomes first order, so we use cyan line to denote the critical points. The s+p coexisting phase generally starts from the s-wave or the p-wave phase, so we use blue lines to denote the emergent points of the s+p coexisting phase on the s-wave phase, and use red curves to denote the emergent points of the s+p coexisting phase on the p-wave phase. When the red or the blue curve becomes dotted, the points denote the emergent points of unstable (part of) s+p solutions. In that case, a first order phase transition occurs. We use the solid yellow curves to denote the critical points of first order phase transitions between the s-wave phases and the p-wave phases, and use the solid green curves to denote the critical points of first order phase transitions between the s-wave phases and the s+p coexisting phases.

In order to describe these phase diagrams more clearly, we define some special values of the dimension of the scalar order to mark vertical coordinates of some critical points. We list their definitions as follows.

\begin{enumerate}
\item $\Delta_{p0}$: The coordinate of the red line at zero temperature.
\item $\Delta_{s0}$: The coordinate of the blue line at zero temperature.
\item $\Delta_{t}$($\Delta_{t+},\Delta_{t-}$): The coordinate of the intersection point of three or four solid lines. For the last three figures, there are two such intersection points, we denote its vertical coordinate value as $\Delta_{t+}$ and $\Delta_{t-}$, respectively, with $\Delta_{t+}>\Delta_{t-}$.
\item $\Delta_{I}$: The coordinate of the contact point of the blue line and the green line.
\item $\Delta_{p min}$, $\Delta_{s min}$: These two denote the coordinates for the lowest points on the red line and blue line respectively. But these two are only useful in  some limited cases, for example, $\Delta_{p min}$ in $b=0.5,0.6$ and $\Delta_{s min}$ in $b=0.3$.
\end{enumerate}

Next  we will discuss   each phase diagram case by case. In each phase diagram, the vertical coordinate is the scaling dimension of the s-wave order parameter, and the horizontal coordinate is the temperature over the fixed value of chemical potential $T/\mu$. In order to explain the phase diagrams more clearly, we can fix the operator dimension to a special value $\Delta=\Delta_{e.g.}$, that means we choose a special horizontal line in the phase diagram, with the vertical coordinate value equal to $\Delta_{e.g.}$. We then force the system to be cooled slowly from a high temperature state in the normal phase region, so the point representing the phase of the system will go from the right side to the left along the line with $\Delta=\Delta_{e.g.}$. We can see how many phase transitions would occur from the phase diagram, and we can also get the critical temperature as well as the order of each phase transition from the phase diagram. Notice that each condensation figure in the previous section shows the case of a horizontal line in one of the phase diagrams.

Now let us begin from the first phase diagram with $b=0.1$. The structure of this phase diagram is quite simple, it is very similar to the phase diagram in the probe limit~\cite{Nie:2013sda}. We can see that in this case $\Delta_{p0}>\Delta_{s0}>\Delta_t$. We can analyze the system in four regions of the parameter $\Delta_{e.g.}$. When $\Delta_{e.g.}<\Delta_t$, the system undergoes only one phase transition from the normal phase to the s-wave phase and stays  in the s-wave phase in the low temperature region. When $\Delta_t<\Delta_{e.g.}<\Delta_{s0}$, there are three phase transitions. The system  will first undergo a p-wave phase transition from the normal phase, then experience a second order phase transition from the p-wave phase to the s+p coexisting phase, and finally meet the final phase transition and change from the s+p coexisting phase to the s-wave phase. The left plot in Figure~\ref{cond-b0102} shows the typical condensation behavior of the system in this region. When $\Delta_{s0}<\Delta_{e.g.}<\Delta_{p0}$, the system will also experience the p-wave phase transition and the phase transition from the p-wave phase to the s+p coexisting phase, but without the phase transition from the s+p coexisting phase to the s-wave phase, the system will finally stay in an s+p coexisting phase in the low temperature region. When $\Delta_{e.g.}>\Delta_{p0}$, there is no s+p coexisting phase, and the the system will only undergo the p-wave phase transition and keep in the p-wave phase in the low temperature region.

The $b=0.2$ phase diagram looks also very similar to the case in the probe limit. The only difference is that the region for the s+p coexisting phase is larger. There are also four regions for the parameter $\Delta$, in these regions, the phase transition properties are qualitatively the same as that in the case $b=0.1$. We plot the condensation behavior of the system with $\Delta_{s0}<\Delta_{e.g.}<\Delta_{p0}$ in the right plot of Figure~\ref{cond-b0102}.

In the $b=0.3$ case, the phase diagram still looks similar to the previous two cases. But it seems that the slope of the blue line changes to be positive in some region. We draw an enlarged version of the phase diagram to show the case clearly. From the enlarged version we can see that the blue line has a minimum value of $\Delta=\Delta_{smin}$ at a non zero temperature, below this temperature the blue line has a negative slope. We have four special values of $\Delta$ with $\Delta_{s min}<\Delta_{s0}<\Delta_t<\Delta_{p0}$, thus we have five regions for $\Delta_{e.g.}$. The most interesting one is the region with $\Delta_{s min}<\Delta_{e.g.}<\Delta_{s0}$, and the condensation behavior for this case is shown in Figure~\ref{cond-b03}. In this region, there would be three phase transitions. The system first goes from the normal phase to the s-wave phase via an s-wave phase transition, then it goes from the s-wave phase to the s+p coexisting phase by the second phase transition. In the third phase transition, the system goes back to the s-wave phase from the s+p coexisting phase. In this case, the behavior of the p-wave order is rather interesting, the condensation value of the p-wave order emerges from zero at some critical temperature. Rather than increasing monotonically,  it decreases after reaching its maximal value and finally vanishes when the system goes back to the s-wave phase. We plot the condensate of the s-wave and  the p-wave orders for this kind of phase transition in Figure~\ref{cond-b03} and call this condensation as an ``n-type'' one. From the phase diagram we can see that the ``n-type''  condensation  occurs because the blue line has a minimum value of $\Delta$ at a non-zero temperature.

In the $b=0.4$ case, the phase diagram is qualitatively the same as that in the case with $b=0.3$, so the above analysis of the $b=0.3$ phase diagram is still valid here. However, it should be pointed out that the lowest point of the blue line seems to be in the dashed part of the line. Because the dashed part is obtained by extrapolation, the lowest point needs to be confirmed by further numerical calculations.

In the $b=0.5$ phase diagram, the blue line always has a positive slope, but the red line has a minimum value $\Delta=\Delta_{p min}$ at a non-zero temperature in this case.  Then there are four special values of $\Delta$ which will divide the phase diagram into five different regions. The interesting part is the region $\Delta_{p min}<\Delta_{e.g.}<\Delta_t$. In this region, the s-wave phase transition occurs first, then the system goes into the s+p coexisting phase via a second phase transition. In the third and forth phase transitions, the system goes into the p-wave phase and then goes back to the s+p coexisting phase. We plot the condensation  for this case in Figure~\ref{cond-b0506}, and we call this kind of condensation as ``u-type''. We can see easily from the phase diagram that the ``u-type'' condensation  can occur because the red line has a minimum value of $\Delta$ at a non-zero temperature.

In the above five phase diagrams, there is always a quadruple intersection point connecting to all the four phases, and all the phase transitions are second order. In the next three phase diagrams, there does not exist any quadruple intersection point, but there are two triple intersection points. And we can see that in this case, some phase transitions become first order.

\begin{figure}
\centering
\includegraphics[width=8cm] {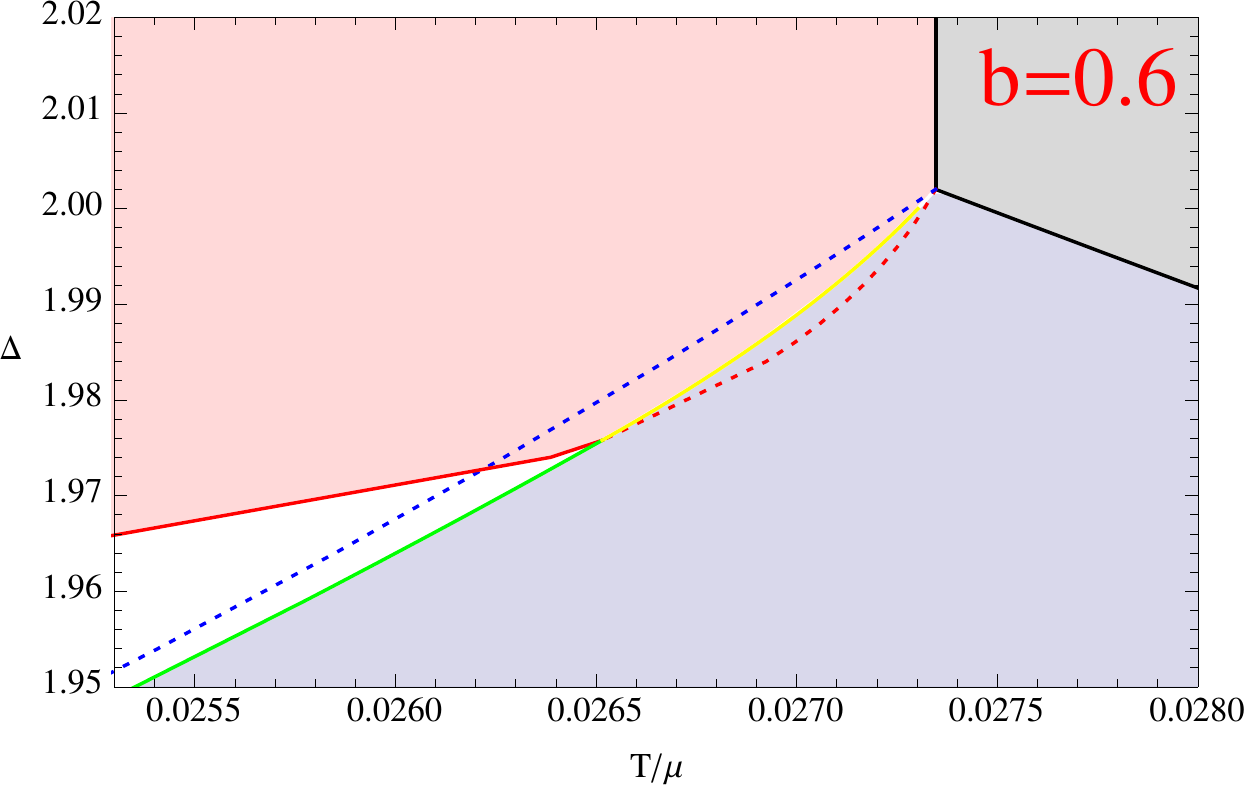}
\caption{\label{PDb06}The enlarged version of the $\Delta-T$ phase diagrams for the case with $b=0.6$.}
\end{figure}

In the phase diagram with $b=0.6$, the p-wave phase transition from the normal phase  is still second order, but in the region $\Delta_I<\Delta_{e.g.}<\Delta_{t-}$, the phase transition from the s-wave phase to the s+p coexisting phase becomes first order. When $\Delta_{t-}<\Delta_{e.g.}<\Delta_{t+}$, the s+p coexisting phase connecting the s-wave phase and p-wave phase still exists, but it always has a higher free energy than the s-wave and p-wave phases. Thus a first order phase transition occurs between the s-wave phase and the p-wave phase. But as we have discussed in the previous section, the unstable s+p coexisting phase is still essential to form the swallow tail shape of the free energy curve for the first order phase transition. In this phase diagram, the red line still has a minimum point as in the $b=0.5$ phase diagram, then the system would also go into the p-wave phase from an s+p coexisting phase and goes back to the s+p coexisting phase later when $\Delta_{p min}<\Delta_{e.g.}<\Delta_{t-}$. We plot the condensation behavior for this case in the right plot of Figure~\ref{cond-b0506}. We draw an enlarged version of this phase diagram in Figure~\ref{PDb06} to show the details near the triple intersection points.

When $b=0.7$, the p-wave phase transition from the normal phase  becomes  first order, so the vertical line separating the p-wave phase and normal phase is marked cyan. In this case, the s+p phase transition from the s-wave phase is still first order in the region $\Delta_I<\Delta_{e.g.}<\Delta_{t-}$, and the s+p coexisting phase connecting the s-wave and p-wave phases is always unstable in the region $\Delta_{t-}<\Delta_{e.g.}<\Delta_{t+}$.

The $b=0.8$ phase diagram is very similar to the one with $b=0.7$, but notice that the green line seems not to intersect with the blue line, which means the phase transition from the s+p coexisting phase to the s-wave phase will be  alway first order. We show the condensate behaviors in Figure~\ref{cond-b08} for three typical regions for $\Delta_{e.g.}$. In Figure~\ref{cond-b08}, the first plot is for $\Delta_{p0}<\Delta_{e.g.}<\Delta_{t+}$, the second plot is for the case $\Delta_{t-}<\Delta_{e.g.}<\Delta_{p0}$, and the third one is for $\Delta_{s0}<\Delta_{e.g.}<\Delta{t_-}$. We also plot the free energy for the three cases in Figure~\ref{FreeE}.

From the above analysis, we can see that the qualitative condensation behaviors at fix $b$ and $\Delta$ can be read off  easily along a horizontal line in each  phase diagram. The different shape of the curves separating the four phases leads to different phase transition behaviors. For example, the ``n-type'' behavior  in Figure~\ref{cond-b03} is due to the existence of  the minimum of the blue line in the $b=0.3$ phase diagram, while the ``u-type'' behavior  in Figure~\ref{cond-b0506} is due to  the existence of  the minimum of the red lines in the $b=0.5$ and $b=0.6$ phase diagrams.

In order to better understand the above phase transition behaviors, we try to find out the possible mechanism behind them. Because the phases with individual s-wave or p-wave order have been well studied, the key issue is to find the underlying law for the emergence of s+p coexisting phase. However, our study is based on numeral solutions, thus we can only give some heuristic argument as follows.

In the holographic superconductor model with single s-wave or p-wave order parameter, the condensate of the order begins when the electric field near the black brane horizon is large enough to support formation of the charged hair. If we lower the temperature with fixed value of the chemical potential in the normal phase, the electric field near the horizon is increasing. Therefore, after the strength of the electric filed is larger than some critical value, the charged order will condense. In the case of our s+p model, the s-wave and p-wave order generally have different critical temperature, therefore the one with higher critical temperature will condense at first. When the first order begins to condense, the condensed order will deplete some charge from the horizon. As a result, the strength of the electric field near horizon is reduced to a smaller value than that of the normal phase at the same temperature. Therefore the condensate of the second order in presence of the first order should occur at a lower temperature than the condensate of the second order on the normal phase (without the first order). We can also confirm this from the phase diagrams in this section.

Another important quantity in the study of the coexisting phase is the free energy. We find that the s+p phase always exists once the free energy curves for the s-wave phase and p-wave phase intersect. It might be possible to get a qualitative formula to predict the existence of the s+p phase, by using some information in free energy. However, this need to be further investigated in future.

We can also see the influence of the back reaction from the phase diagrams. The most obvious one is that the region for the s+p coexisting phase is enlarged by the back reaction when the back reaction strength $b$ is not too large. This is in accordance with the intuition that the back reaction on the metric adds an additional attraction  between the s-wave and p-wave orders. This also happens in the s+s system in Ref.~\cite{Cai:2013wma}.

However, we here have found that  for a very large value of $b$, the region for the s+p coexisting phase is reduced. This is mainly caused by the low critical temperatures of the s-wave and p-wave phase transitions in the strong back-reaction cases. We should also notice that the s+p solution near the triple intersection points becomes unstable in the strong back reaction case, this also makes the region for the s+p coexisting phase smaller.

In the parameter region we have tested, the s-wave phase transition from the normal phase is always second order, but the p-wave phase transition becomes first order when the back reaction is large enough. The critical value of the back reaction at which the p-wave phase transition becomes first order can be read off  from Ref.~\cite{Herzog:2014tpa}, and the critical value is $b_c=0.62$. In this work, we  see that the phase transition from the s-wave phase to the s+p coexisting phase can also be first order when the back reaction is large. And even when both the p-wave and s-wave phase transitions are still second order, the s+p coexisting phase transition can  be first order, as exhibited in the phase diagram of $b=0.6$.

\section{An alternative setup of the holographic s+p model}\label{sect:newSetup}
Before we conclude our work, we give another setup for a holographic s+p model in this section. From this new setup, we can get the same equations of motion as those presented in  Sec.~\ref{sect:setup}. Thus we can get the same condensation behavior and the same phase diagrams as in the previous sections.

It is well known that the origin of superconductivity involves the formation of a quantum condensate state by pairing conduction electrons.  Nevertheless, from a symmetry point of view, the phenomenon of superconductivity can be thought of as the spontaneously breaking of U(1) gauge symmetry. In the setup discussed above, the p-wave and s-wave order parameters are dual to the gauge field pointing in the third direction inside the SU(2), i.e.,  $A^3_\mu$ and the third component of the scalar triplet $\Psi^3$, respectively. By fixing a gauge, we choose the SU(2) gauge field in the first direction as the U(1) gauge symmetry which we call $U(1)_1$. A non-vanishing value of the time component of $U(1)_1$ gauge field induces a chemical potential on the dual theory. The non-trivial profile of the gauge field $A^3_x$ would induce a vacuum expectation value of the dual operator with no source and thus breaks the $U(1)_1$ symmetry spontaneously. Similar story is also true for the scalar field $\Psi^3$.

However, there are still some limitations in the setup discussed above. One can change the scaling dimension of the dual scalar operator for the s-wave order by adjusting the mass square of $\Psi^a$. But dual to the constraint from the SU(2) gauge symmetry, the dimension of the p-wave order has to be fixed to be $\Delta_p=d-1$ where $d$ is the spatial dimension in the bulk. Furthermore, the SU(2) gauge invariance also makes the ``charge" of the p-wave order equal to the one of the s-wave order. Therefore in our previous setup, we only have two free parameters $b$ and $\Delta_s=\Delta$ at hand. In principle, the dimension and the charge for the p-wave order can also have different values, so we also need to find the way to vary these parameters in the holographic setup. One way is to use the holographic p-wave model~\cite{Cai:2013pda} in terms of a charged complex vector field. In this new setup, we can use a complex scalar and a complex vector minimally coupled with a $U(1)$ gauge field as the matter part of the holographic action. The matter part of the action can be written as
\begin{eqnarray}
S_M'=\int d^{d+1}x \sqrt{-g}\Big(&-\frac{1}{4}F_{\mu\nu}F^{\mu\nu}-\frac{1}{2} \rho^\dagger_{\mu\nu}\rho^{\mu\nu}-m_p^2 \rho^\dagger_\mu \rho^\mu \nonumber \\\label{spnew}&
-\tilde{D}_\mu^\dagger \Psi \tilde{D}^\mu \Psi-m_s^2 \Psi\Psi^\dagger\Big).
\end{eqnarray}
where the superscript ``$^\dagger$" means complex conjugate, $F_{\mu\nu}=\nabla_\mu A_\nu-\nabla_\nu A_\mu$ is the field strength for the U(1) gauge field. Here $\Psi$ is now a complex scalar charged under the U(1) gauge field with $\tilde{D}_\mu =\nabla_\mu-iq_s A_\mu$, where $q_s$ is the U(1) charge of $\Psi$. $\rho_\mu$ is a complex vector field charged under the same U(1) field but with a different charge $q_p$. The field strength of $\rho_\mu$ is $\rho_{\mu\nu}=\bar{D}_\mu \rho_\nu-\bar{D}_\nu \rho_\mu$ with covariant derivative $\bar{D}_\mu =\nabla_\mu -iq_p A_\mu$. The same equations of motion from the previous model under the ansatz~\eqref{matterAnsatz} can also be obtained  in the particular case with $q_p=q_s=g_c$ and $m_p=0$ in this new model under the following ansatz
\begin{eqnarray}\label{newmatterAnsatz}
\Psi=\Psi(r),~A_t=\phi(r),~\rho_x=\Psi_x(r),
\end{eqnarray}
with all other matter field components being set to zero. Thus all results in the previous sections can also be recognized as the $q_s=q_p=g_c$ and $m_p=0$ case in the new s+p model in this section.

It is worth pointing out that although those two s+p models show same aspects in particular cases, they are distinct from each other. For example, the transport coefficients might exhibit essential differences~\cite{Herzog:2014tpa}. And with this new holographic s+p model, we can consider the effect of different charges of s-wave and p-wave orders. Furthermore, we can also introduce the mass term of $\rho_\mu$ to deform the scaling dimension of the p-wave vector operator.  So we can study the s+p system holographically in a more complete parameter space. It has been shown that depending on $q_p$ and $m_p^2$, only the p-wave part of action~\eqref{spnew} can exhibit a rich phase structure~\cite{Cai:2013aca,Cai:2013kaa,Cai:2014ija}. Thus the new model~\eqref{spnew} is expected to show more fruitful phase behaviors~\cite{Arean:2010zw,Wu:2014bba} in a larger parameter space. This will be left for further study.

\section{\bf Conclusions and discussions}\label{sect:conclusion}

In this paper, we have continued our study on the holographic s+p model with  a scalar triplet charged under an SU(2) gauge field in the bulk~\cite{Nie:2013sda},  by  considering the back reaction of the matter fields on the background black brane geometry. We have shown that  the model exhibits a rich phase structure depending on the model parameters. Taking eight different values of the back reaction parameter $b$ from $0.1$ to $0.8$, we have plotted the condensate 
behaviors of the s-wave and p-wave orders and built corresponding phase diagrams in terms of the dimension of the s-wave order and the temperature of the system. We  have also given  another setup for a holographic s+p model, which can lead to the same equation of motion as  the model proposed in~\cite{Nie:2013sda} with special value of model parameters.

In this holographic s+p superconductor model, we have discovered  some new condensate behaviors such as the ``n-type'' and ``u-type'', where the condensations for the p-wave or s-wave orders are  non-monotonic. These interesting behaviors can only occur when the blue or red curve in the phase diagram has a minimum at non-zero temperature. These non-monotonic condensation behaviors are also known as ``reentrant"\footnote{We would like to thank Professor Jan Zaanen for reminding us this issue.} phase transitions in condensed matter physics, and would be very useful in future studies. For example, we can solve the time evolution problem of a quenched initial state in the ``n-type'' to study the non-equilibrium properties in systems with multi-orders and compare the results to that in the system with only the s-wave order.

We also have found that the s+p solution might be totally unstable when the back reaction is strong enough, and a first order phase transition between the s-wave and p-wave phases occurs at that time. Although the s+p solution in that case is totally unstable, it is still essential in forming the swallow tail in free energy curves for the first order phase transition. According to this property, if we assume the swallow tail of free energy curve as a universal feature for the first order phase transitions, we can explain why the s+p solution can always exist when the free energy curves for the s-wave and p-wave phases have an intersection point.

The eight phase diagrams give a unified description for the various condensation behaviors in Sec.~\ref{sect:setup}. Each phase condensation behavior at fixed $\Delta$ and $b$ can be read from a horizontal line in the phase diagram. We can also get a concrete understanding on the effect of the back reaction on our holographic s+p model from the resulting phase diagrams. We can see that the back reaction of the matter sector on the black brane background  generally enlarges the region for the s+p phases when the back reaction is not too large, which is similar to the case of the s+s system~\cite{Cai:2013wma}. But in the very strong back reaction case, the region for the s+p phase becomes smaller. This is mainly caused by the decreasing of the critical temperatures for the s-wave and p-wave phase transitions when the back reaction becomes strong. The fact that the s+p phase becomes unstable near the triple intersection points also contributes to the decreasing of the region for the s+p phase in the phase diagram.

In the strong back reaction case, the p-wave phase transition from the normal phase  becomes first order. With the back reaction getting stronger, there is a growing parameter space where the transition from the s-wave phase to the s+p phase is  first order. The first order phase transition from the s-wave phase to the s+p phase even exists when both the s-wave and p-wave phase transitions are still second order.

There are many further directions to explore based on this work. With the new setup proposed in section \ref{sect:newSetup}, we can study the s+p system more completely with various charges and dimensions of the p-wave order. Since our s+p system has shown many new phase transition behaviors, one can use them to test some universal properties in holographic systems. In addition, these new  s+p models  provide  the setups in which we can study  time dependent or inhomogeneous solutions in order to see the non-equilibrium physics or the lattice effects. Finally if the swallow tail shape of the free energy curve is indeed a universal property for the first order phase transitions, it can be used to predict new phases(solutions) near the intersection point of two different phases.

%=========================================================================
\begin{acknowledgement}
\section*{Acknowledgements}
ZYN would like to thank Elena Mafalda Caceres, Jerome P. Gauntlett, Qi-Yuan Pan, Julian Sonner, Run-Qiu Yang, Xiao-Quan Yu, Jan Zaanen for useful discussions, and thank the organizers of ``Conference on Non-Equilibrium Phenomena in Condensed Matter and String Theory" and the organizers of ``Quantum Gravity, Black Holes and Strings" for their hospitality. XG was supported in part by NSF grant PHY-1417337. LL was supported in part by European Union's Seventh Framework Programme under grant agreements (FP7-REGPOT-2012-2013-1) no 316165, the EU-Greece program ``Thales" MIS 375734 and was also co-financed by the European Union (European Social Fund, ESF) and Greek national funds through the Operational Program ``Education and Lifelong Learning" of the National Strategic Reference Framework (NSRF) under ``Funding of proposals that have received a positive evaluation in the 3rd and 4th Call of ERC Grant Schemes". This work was also supported in part by the Open Project Program of State Key Laboratory of Theoretical Physics, Institute of Theoretical Physics, Chinese Academy of Sciences, China (No.Y5KF161CJ1), and in part by Shanghai Key Laboratory of Particle Physics and Cosmology under grant No.11DZ2230700, in part by two talent-development funds from Kunming University of Science and Technology under Grant Nos. KKZ3201307020 and 

KKSY201307037, and in part by the National Natural Science Foundation of China under Grant Nos.  11035008, 11247017, 11375247, 11447131 and 11491240167.
\end{acknowledgement}
%=========================================================================

%%%%%%%%%%%%%%%
%%%%%%%%%%%%%%%
%%%%%%%%%%%%%%%
%%%%%%%%%%%%%%%
%%%%%%%%%%%%%%%
%%%%%%%%%%%%%%%

\end{document}